\documentclass[
twocolumn,
english,
superscriptaddress,
amsmath,
floatfix,
longbibliography,
floats,
prl,
]{revtex4-2}

\usepackage[colorlinks=true,urlcolor=blue,citecolor=blue,linkcolor=blue]{hyperref} 
\usepackage[T1]{fontenc}
\usepackage[utf8]{inputenc}
\usepackage{amssymb}
\usepackage{graphicx}
\usepackage{amsmath,color}
\usepackage{mathrsfs}
\usepackage{float}
\usepackage{amsmath,bm}

\usepackage{indentfirst}
\usepackage{txfonts}
\usepackage[normalem]{ulem}
\usepackage[table]{xcolor}
\usepackage{array,ragged2e}
\usepackage{grffile}
\usepackage{multirow}
\usepackage{algpseudocode}
\usepackage{epstopdf}

\usepackage[caption=false]{subfig}

\usepackage{graphicx,xcolor} 

\usepackage{comment}

\makeatletter
\newcommand{\bra}[1]{\langle #1 |}
\newcommand{\ket}[1]{| #1 \rangle}

\makeatother
\usepackage{babel}

\begin{document}

\hyphenpenalty=5000
\tolerance=1000

\title{Tensor Network Computations That Capture Strict Variationality, Volume Law Behavior, and the Efficient Representation of Neural Network States}

\author{Wen-Yuan Liu}
\thanks{These authors contributed equally to this work.}
\affiliation{Division of Chemistry and Chemical Engineering, California Institute of Technology, Pasadena, California 
91125, USA}

\author{Si-Jing Du}
\thanks{These authors contributed equally to this work.}
\affiliation{Division of Engineering and Applied Science, California Institute of Technology, Pasadena, California 91125, USA}

\author{Ruojing Peng}
\affiliation{Division of Chemistry and Chemical Engineering, California Institute of Technology, Pasadena, California 91125, USA}
 \author{Johnnie Gray}
\affiliation{Division of Chemistry and Chemical Engineering, California Institute of Technology, Pasadena, California 91125, USA}
\author{Garnet Kin-Lic Chan}
\email{gkc1000@gmail.com}
\affiliation{Division of Chemistry and Chemical Engineering, California Institute of Technology, Pasadena, California 91125, USA}

\date{\today }
\begin{abstract}

We introduce a change of perspective on tensor network states that is defined by the computational graph of the contraction of an amplitude. The resulting class of states, which we refer to as tensor network functions, inherit the conceptual advantages of tensor network states while removing computational restrictions arising from the need to converge approximate contractions. We use tensor network functions to compute strict variational estimates of the energy on loopy graphs, analyze their expressive power for ground-states, show that we can capture aspects of volume law time evolution, and provide a mapping of general feed-forward neural nets onto efficient tensor network functions. Our work expands the realm of computable tensor networks to ones where accurate contraction methods are not available, and opens up new avenues to use tensor networks.

\end{abstract}
\maketitle

\noindent \paragraph{\it Introduction.} Simulating and representing quantum many-body systems is a central task of physics. For this, tensor network states have emerged as a fundamental language and numerical tool~\cite{white1992,verstraete2008,schollw2011,chan2011the,tnsRMP2021,xiang2023density}, with applications across diverse areas such as condensed matter physics~\cite{gu2008,schuch2011,corboz2014competing,liao2017,zheng2017stripe,liu2022emergence}, statistical physics~\cite{levin2007,xie2009second,evenbly2015tensor,yang2017loop}, quantum field theory~\cite{verstraete2010,jutho2013,taglia2014}, and quantum information theory~\cite{vidal2003efficient,markov2008simulating,arad2010}, as well as adjacent disciplines such as quantum chemistry and machine learning~\cite{bogu2012,naka2013,miles2016,han2018}.

Tensor networks represent the amplitude of a quantum state on an $L$ site lattice, denoted $\langle {\bf n}|\Psi\rangle \equiv \Psi({\bf n}) \equiv \Psi(n_1, n_2, \ldots, n_L)$, as the contraction of a set of tensors. As a simple example, we consider $L$ tensors connected in a graph (Fig.~\ref{fig:tns}a), each associated with a lattice site Hilbert space; this is known as a projected entangled pair state (PEPS)~\cite{PEPS2004}. Each tensor contains $dD^m$ elements ($m$ is the number of bonds which connect tensors to adjacent tensors): $d$ is the local Hilbert space dimension and $D$ controls the expressivity.  For tree graphs, the resulting contraction over the shared tensor bonds can be computed exactly in $O(L)\mathrm{poly}(D)$ time~\cite{treeTensor}. For general graphs, however, both  $\langle {\bf n} | \Psi\rangle$ and $\langle \Psi |\hat{O}|\Psi\rangle$ cannot be exactly contracted efficiently: the cost is exponential in the tree-width of the graph~\cite{markov2008simulating}. Approximate contraction algorithms have thus been introduced whereby during the contraction process, the large intermediate tensors that are formed are compressed, via singular value decomposition (SVD), to a controlled size, $O(\mathrm{poly}(\chi))$, where $\chi$ is termed the auxiliary bond dimension~\cite{verstraete2008,ran2020tensor}. Only in the limit of $\chi\to \infty$ is the contraction exact. The resulting approximate contraction  has a cost $O(L)\mathrm{poly}(D)\mathrm{poly}(\chi)$, with an error with respect to the exact contraction controlled by $\chi$.

From the perspective of approximate contraction, tensor network computations should seek to be converged with respect to $\chi$, otherwise the results are not meaningful. For example, for a Hamiltonian $\hat{H}$, if $\langle \Psi | \hat{H}|\Psi\rangle / \langle \Psi|\Psi\rangle$ is computed by approximate contraction, with insufficient $\chi$ the energy may violate the variational theorem. The requirement that computations should use a sufficiently large $\chi$ (such that the evaluation of $\langle \Psi | \hat{O}|\Psi\rangle$ or $\langle {\bf n} |\Psi\rangle$ is no longer changing significantly with $\chi$) places restrictions on the types of graphs that can be used to define the tensor network. For example, in many-body physics simulations, it is believed that only highly structured graphs with limited entanglement, such  as (near-)area law states, support an efficient contraction scheme. Because of the latter requirement, it has been argued that, in practical use, tensor networks cannot replicate the expressivity of other flexible types of many-body ansatz, such as neural network quantum states~\cite{changlani2009approximating,deng2017quantum,chen2018equi,glasser2018,levine2019quantum,pastori2019,nns2022}.

In the current work, we observe that the above limitations on tensor network computations arise due to an assumption, namely that the tensor network ansatz is only meaningfully defined in the exact contraction limit. We can, however, adopt an alternative perspective, namely, to consider the approximate contraction scheme for any $\chi$ as itself part of the tensor network ansatz for the amplitude $\Psi({\bf n})$. 
This view is natural and in some ways obvious when using tensor networks in variational Monte Carlo (VMC) calculations~\cite{sandvik2007,schuch2008,wang2011,liu2017,liu2021}, but we will argue that this is not a manifestation of tensor networks in a particular numerical algorithm, but rather a change of perspective that alters and broadens the definition and utility of tensor networks. We refer to the larger set of parameterized many-body states that arise from ``consistent'' contraction schemes (to be defined later)  as \emph{tensor network functions}. This wider class of states removes many restrictions that are assumed to apply to tensor networks. We demonstrate that tensor network functions provide strictly variational estimates of the energy, allow for practical tensor network computations on graphs that are highly challenging to contract accurately, including those that describe volume law physics, and expand the set of states that can be efficiently represented by tensor networks to cover other classes of ansatz, such as neural network quantum states.

\begin{figure}[htbp]
\includegraphics[width=\columnwidth]{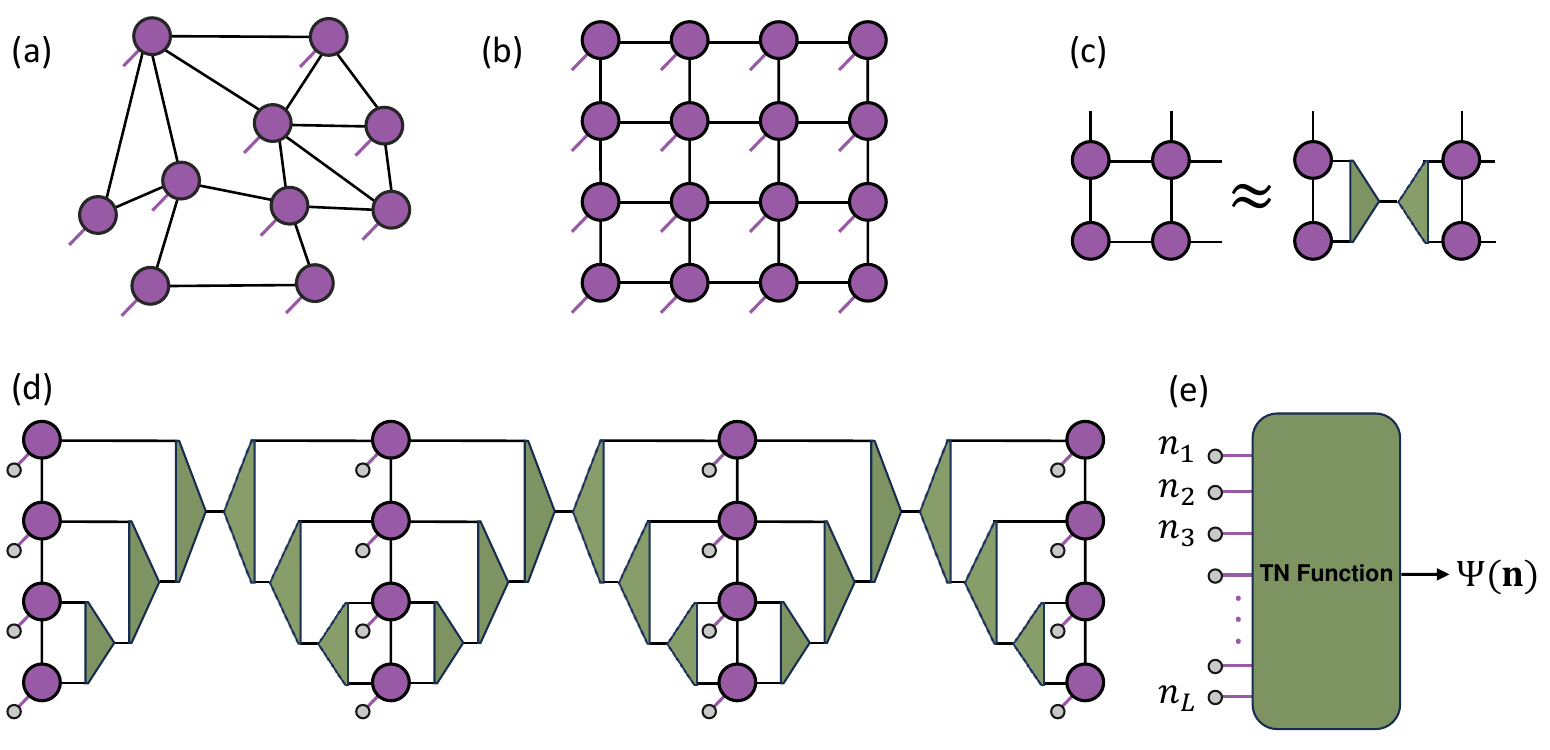}
\captionsetup{justification=justified,singlelinecheck=false}
\caption{\label{fig:tns} (a) PEPS on a general graph with $L$ site tensors. (b) PEPS on a $4\times4$ square lattice. (c) Insertion of a pair of isometries for approximate tensor contractions; the isometries can be obtained with standard tensor network techniques by SVD~\cite{verstraete2008,schollw2011}. (d) Approximate PEPS contractions by inserting isometries for a given configuration $\ket{{\bf n}}$ (gray circles). (e) For a given input vector $\ket{{\bf n}}=(n_1,n_2,..., n_L)$, a tensor network function outputs a unique value $\Psi({\bf n})$. }
\end{figure}

\noindent \paragraph{\it Tensor network functions.} A tensor network function (TNF) is a consistent contraction of the tensor network representation of the amplitude $\Psi({\bf n})$. To illustrate the basic idea, we first consider a standard PEPS for a two-dimensional lattice, illustrated in Fig.~\ref{fig:tns}b. Because the graph has loops, the PEPS amplitude cannot be computed efficiently according to Fig.~\ref{fig:tns}b. Instead, in practice, the  amplitude is defined by a different tensor network illustrated in Fig.~\ref{fig:tns}d. In this figure, isometries (triangles) with bond dimension $\chi$ have been inserted, where the values of the isometries are determined by an SVD of the contraction of neighbouring tensors (see Fig.~\ref{fig:tns}c)~\cite{verstraete2008,schollw2011}. The final amplitude is obtained by contracting the tensors and isometries together. The pattern of isometries used here corresponds to a specific approximate contraction method (boundary contraction~\cite{PEPS2004,verstraete2008}) and other choices can be made~\cite{ran2020tensor,gray2024hyperoptimized}, but the essential purpose is to make the contraction of the modified graph containing the original tensors and the isometries efficient.

Note, however, that although the structure of the tensor network with isometries in Fig.~\ref{fig:tns}d now appears pseudo-one-dimensional, this does not mean that its entanglement structure (for $\chi > 1$) is reduced to that of a matrix product state (MPS). This is because the isometries are themselves functions of the configuration $\ket{{\bf n}}$, i.e. different values of the isometries are obtained, via SVD, for different configurations. 
However, we choose to keep the same \emph{positions} of the isometries in the graph for any configuration (even though the entries are different); we refer to this as using \emph{fixed} (position) isometries, which defines a consistent contraction. Because of the configuration dependence of the isometry entries, the graph with isometries no longer represents a multi-linear ansatz like the original tensor network state; it is only a multi-linear ansatz in the limit that $\chi \to \infty$ (and the isometries become identities). Instead, it defines a \emph{tensor network function}: for any given $(D, \chi)$, any $\ket{{\bf n}}$ is mapped to a single value, the amplitude, through an efficient contraction. The efficient TNF representation of $\Psi({\bf n})$ does not guarantee deterministic efficiency of computing $\langle \Psi|\hat{O}|\Psi\rangle$; but such expectations may be evaluated stochastically by sampling with efficient sample complexity.

We emphasize that here we are defining a new class of state i.e. the TNF, whose amplitudes are exactly given by a certain contraction graph of a tensor network. The essential differences between TNF and conventional tensor network computations are further illustrated in the following examples, and discussed also in Section-1 in SM~\cite{SM}.

\noindent \paragraph{\it Tensor network functions and approximate contraction.} VMC algorithms also compute $\Psi({\bf n})$ via approximate contraction (for a brief review of tensor network VMC (TN-VMC) algorithms, see SM~\cite{SM} or Refs.~\cite{sandvik2007,schuch2008,wang2011,liu2017,liu2021}) and it is instructive to consider the differences between the standard TN-VMC approximation of $\Psi({\bf n})$ and the definition of a TNF above. 
There are two seemingly small, but significant, differences (i) in TN-VMC, $\chi$ is viewed as a parameter (typically $\chi = \mathrm{const} \times D$, with $\mathrm{const} > 1$) where $\chi$ is increased to convergence for a fixed $D$, and (ii) the positions of the isometries are not fixed but dependent on the amplitude that is being computed (we refer to this as \emph{dynamic} isometries), in order to allow for the reuse of computations when computing amplitudes of configurations which differ by only a few substitutions~\cite{liu2021}.  See SM for details~\cite{SM}.

The most important distinction from TNF is that the use of {dynamic} isometries in TN-VMC \emph{does not lead to a consistent function}. In other words, given the same configuration $\ket{{\bf n}}$, $\Psi({\bf n})$ needs not be the same, as it depends on the prior configuration and which components of the tensor network contraction are being reused. Only in the limit of $\chi \to \infty$ are the amplitudes consistent and the function defined, which underlies the usual need to converge $\chi$ in TN-VMC calculations. In contrast, it is clear that with using fixed isometries, the corresponding TNF is consistent for any $D$ and any $\chi$. See SM for details~\cite{SM}.

\begin{figure}[htb]
\centering
\includegraphics[width=\columnwidth]{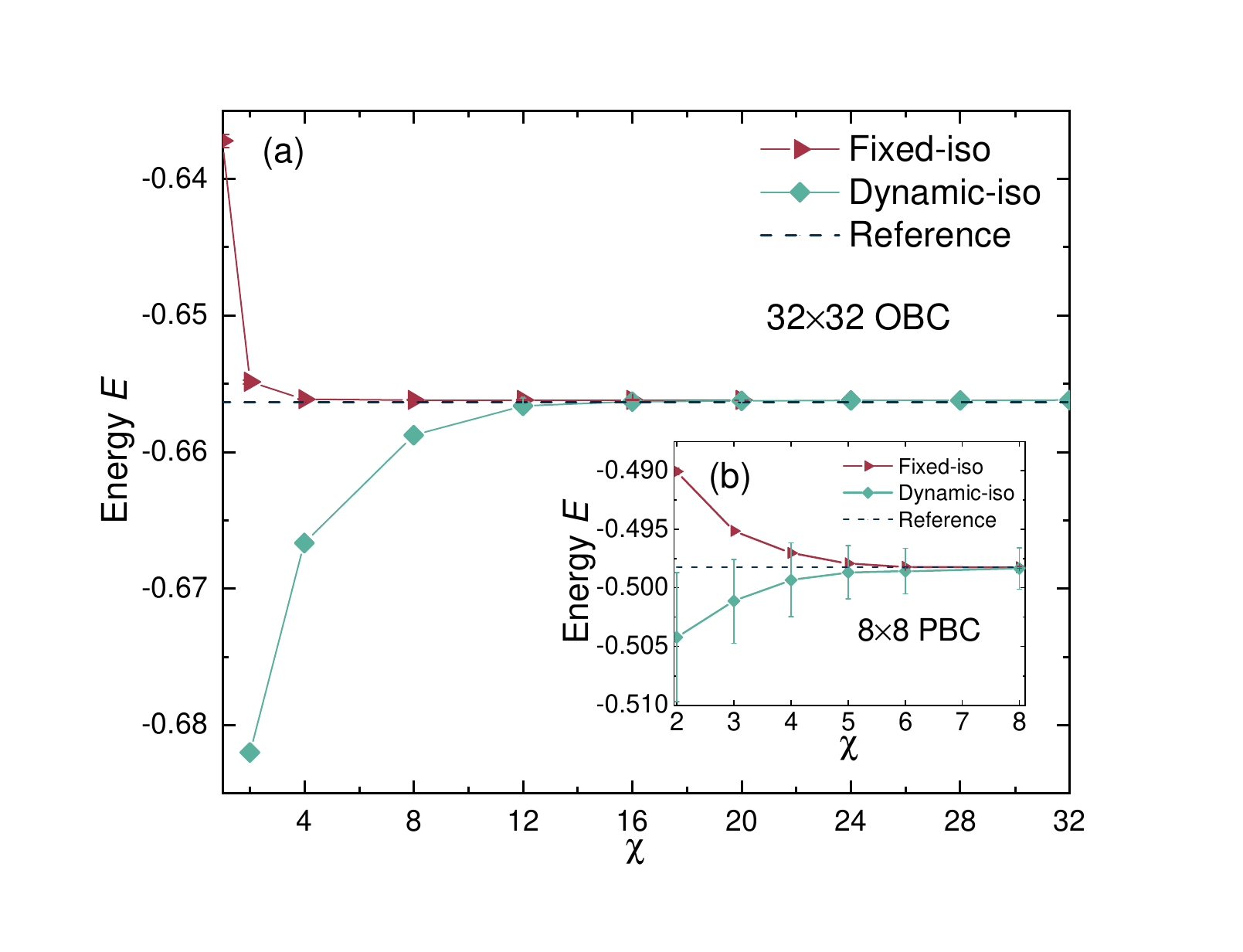}
\caption{ For a given PEPS, energies (per site) obtained from fixed (tensor network function) and dynamic (standard VMC) isometry contractions. (a) OBC square-lattice $32\times 32$ spin-1/2 Heisenberg model with PEPS $D=8$, and 15000 Monte Carlo sweeps are used for each data point. The dashed line denotes the numerically exact energy~\cite{liu2021}. (b) PBC square-lattice $8\times 8$ spin-1/2 frustrated $J_1-J_2$ model at $J_2/J_1=0.5$ with PEPS $D=5$, and 5000 Monte Carlo sweeps are used for each data point. The dashed line denotes the energy from fixed-isometry contraction with $\chi=8$ using $D=5$.}    
\label{fig:vmc}
\end{figure}

\paragraph{\it Tensor network functions and variational energies.} As an immediate consequence of the use of fixed isometries to define a TNF, we can always obtain a variational energy in Monte Carlo energy evaluation (up to statistical error).

In Fig.~\ref{fig:vmc}(a) we show the energy of an optimized PEPS ($D=8$, various $\chi$) TNF for the ground-state of the open-boundary condition (OBC) spin-1/2 square-lattice $32 \times 32$ Heisenberg model defined using the boundary contraction computational graph, see SM, compared to that of a PEPS in standard TN-VMC using dynamic isometries, where $\chi$ controls the error of approximate contraction~\cite{liu2021}.
We see immediately that the PEPS TNF energy is variational for all $\chi$. In contrast,  viewing the PEPS amplitude as an (inconsistent) approximate contraction as in standard VMC leads to a non-variational energy, where the non-variationality is typically regarded as the ``contraction error''. This behaviour is even more clear for a lattice that is usually considered more difficult to contract. Using the square-lattice $J_1-J_2$ model with periodic boundary conditions (PBC) as an example,  in Fig.~\ref{fig:vmc}(b) we show the same comparison between TNF and standard TN-VMC. In addition to the variational versus non-variational nature of the energies, we also see a large variance of the TN-VMC energy arising from the inconsistent definition of amplitudes. This leads to substantial difficulties in the stochastic optimization of the ansatz. Overall, we see that lack of variationality is not intrinsic to approximate contraction: it results from the perspective on the definition of the ansatz.

\paragraph{\it Expressivity of tensor network functions with small $\chi$.}

\begin{figure}[htb]
\includegraphics[width=0.99\columnwidth]{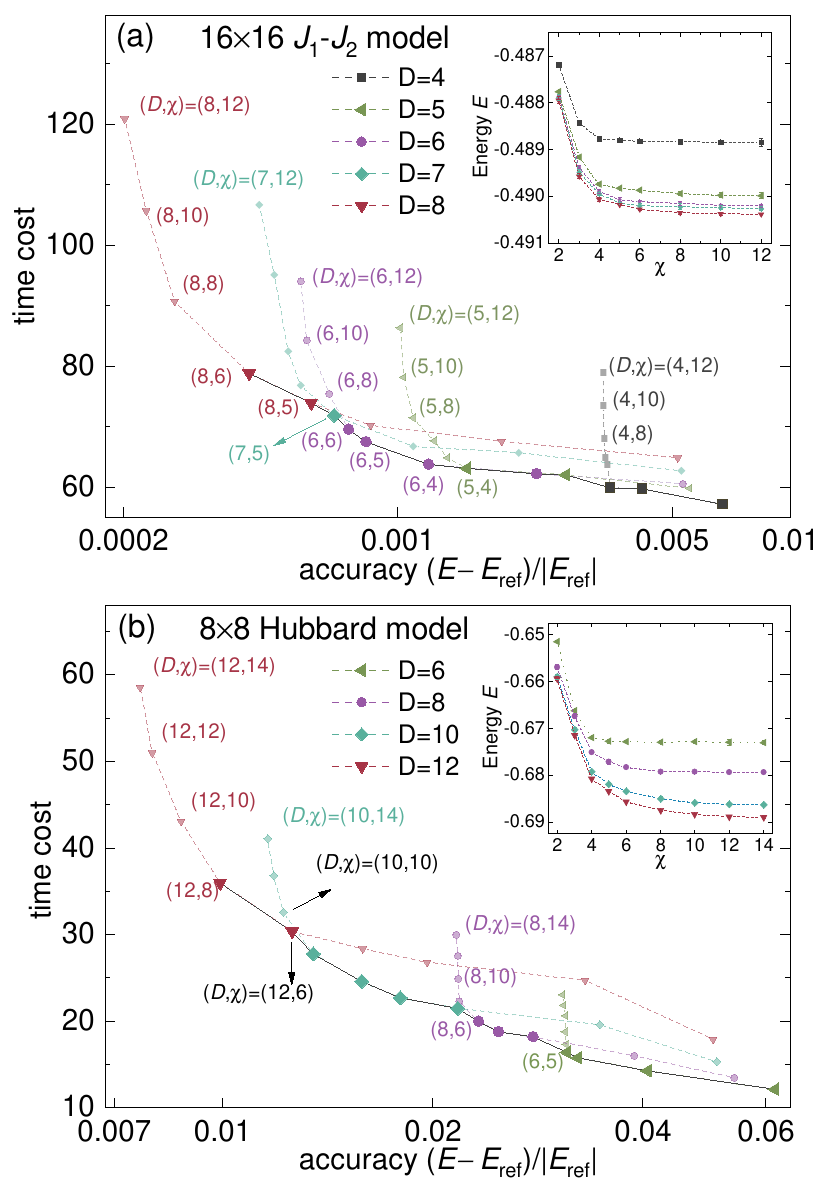}
\caption{ Time cost (in seconds) versus accuracy of the tensor network function $|\Psi(D,\chi)\rangle$, for the $J_1-J_2$ model on the $16\times 16$ lattice at $J_2/J_1=0.5$ (a) and Fermi-Hubbard model on the $8\times8$ lattice at $U/t=8$ with hole doping $n_h=1/8$ (b).  Reference energies $E_{\rm ref}$ taken from PEPS $D=10$ ($J_1-J_2$, $E=-0.490478(4)$) and DMRG ($D=10000$ with SU(2) symmetry, Hubbard model, $E=-0.694391$). 
The time cost is for computing $\langle {\bf n} |\Psi(D,\chi)\rangle$ for a given $| {\bf n} \rangle $ on a single core.  Highlighted symbols (connected by a black line) indicate the optimal $(D,\chi)$ pairs that have smallest time cost for a given accuracy. Here $D=4-8$ and $\chi=2\sim 12$ is used for the $J_1-J_2$ model and $D=6-12$ and $\chi=2\sim 14$ for the Hubbard model. The insets present the energy variation of the TNF $|\Psi(D,\chi)\rangle$ with $D$ and $\chi$.} 
\label{fig:accuracy_time} 
\end{figure}

In general, a TNF is parameterized by two bond dimension parameters, $D$ and $\chi$. We already know that as $\chi$ becomes large, the TNF reduces to the standard tensor network state from which it is derived, but TNFs are well defined also in the limit of small $\chi$ (e.g. $\chi < D$) and it is important to ask whether they are expressive in this limit. 

In Figs.~\ref{fig:accuracy_time}(a) and (b), we show results from approximating the ground-state of a spin-$1/2$ square-lattice  $16 \times 16$  $J_1-J_2$ Heisenberg model for $J_2/J_1=0.5$, and an $8\times 8$ Hubbard model with $U/t=8$ and $1/8$ hole doping, 
using bosonic and fermionic PEPS TNFs for OBC systems~\cite{PEPS2004,liu2022,fPEPS2010,lee2023EQA}, respectively. We minimize the energy for many different $(D, \chi)$ combinations using gradient descent, and we plot the computational cost of the amplitude against the obtained accuracy. For a specified relative accuracy, we can then search for the $(D, \chi)$ pair with the lowest computational cost. We see that contrary to the standard situation with tensor network states where one uses $\chi > D$, the most powerful TNFs for a given computational cost have $\chi < D$. For example, for $\sim 0.01$ relative accuracy in the Hubbard model, using $(D, \chi) = (12, 6)$ yields the same expressivity as $(D, \chi) = (10, 10)$, but with lower cost. In no case is it computationally more efficient to use $\chi > D$. 
The insets of Fig.~\ref{fig:accuracy_time} presenting the minimized energy for every ansatz $|\Psi(D,\chi)\rangle$, show that the most rapid improvement in energy is obtained at small $\chi$, rather than saturating $\chi$ for a given $D$, confirming the above conclusion.

\begin{figure*}[tbp]
\centering
 \includegraphics[width=0.3\textwidth]{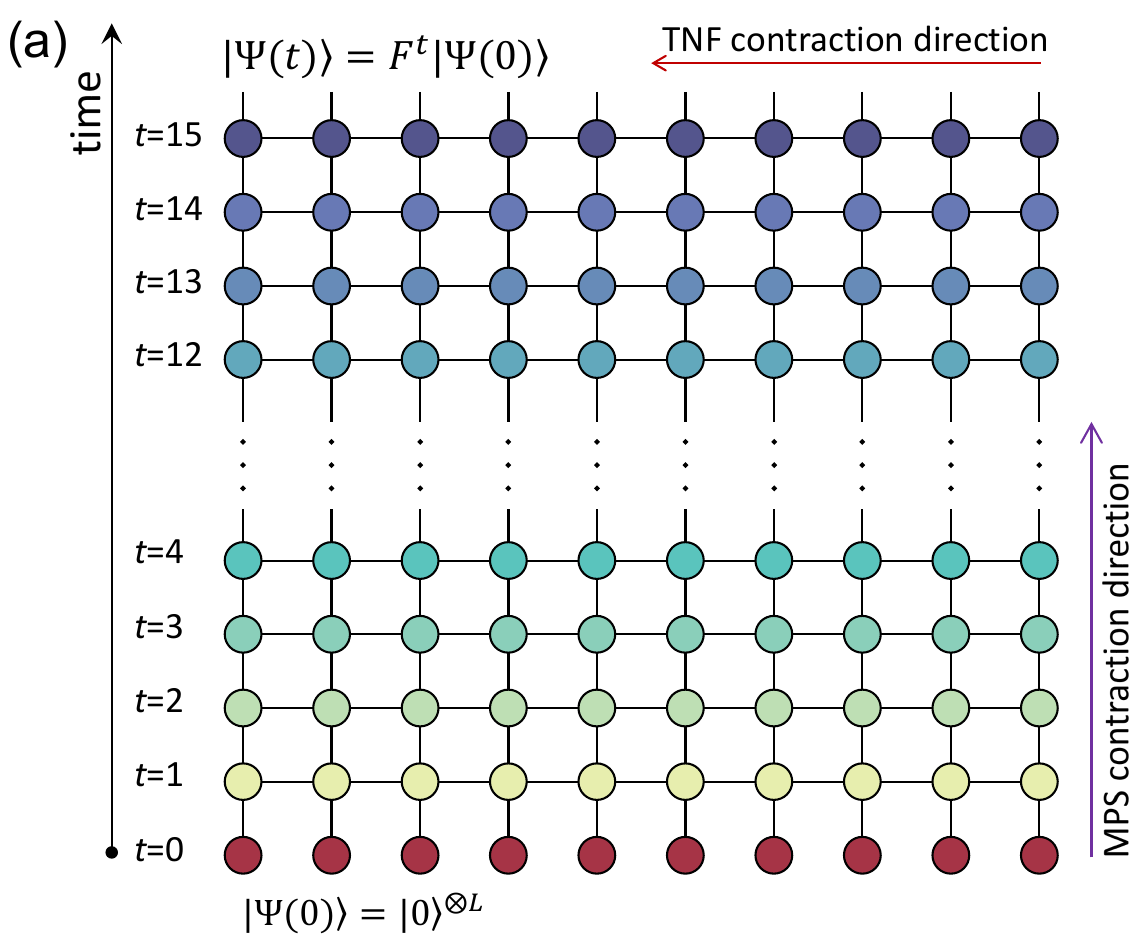}
\includegraphics[width=0.68\textwidth]{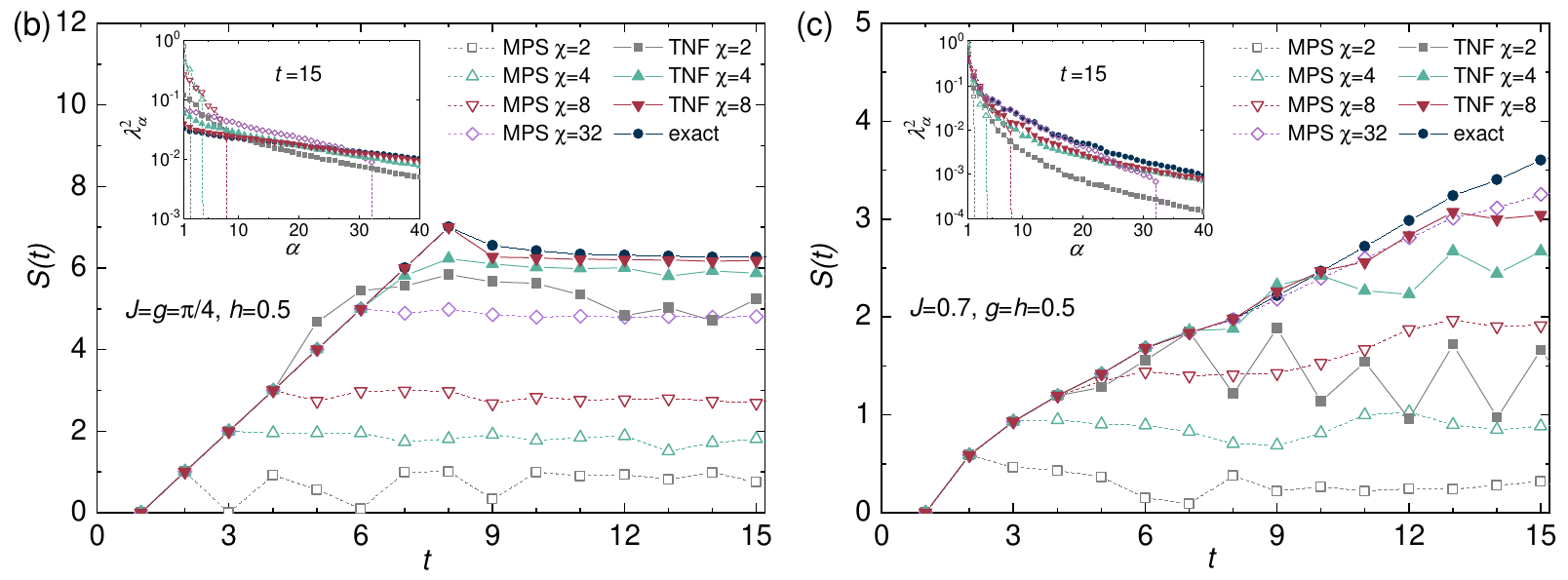}
\caption{\label{fig:volumelaw}{ (a) (1+1)D TN representation of the  Floquet dynamics of an $L$-site kicked Ising chain, starting from a product state. TNF contraction direction is either along the spatial direction or inverse time directon (see inverse direction results in SM~\cite{SM}). Conventional MPS-MPO contraction is along the time direction. (b) and (c) show the dynamics of the half-system bipartite von Neumann entropy $S(t)$ of a 14-site chain at $(J,h,g)=(\pi/4,\pi/4,0.5)$ and  $(0.7,0.5,0.5)$, respectively. Insets of (b) and (c) present the entanglement spectrum for the 40 largest eigenvalues of the half-chain reduced density matrix at $t=15$.}}
\end{figure*}

\paragraph{\it Tensor network functions and volume laws.} As an example of a new kind of application that we can consider with TNFs, we consider the case where the tensor network graph corresponds to chaotic time evolution leading to a volume law state. Then the standard view  is that approximate contraction should be performed with $\chi \sim \exp(\mathrm{const} \times t)$ (where $t$ is evolution time) to capture the volume law. 
However, if we think of the TNF formed by a random circuit, we might expect that the set of amplitudes, defined using fixed position but configuration dependent and essentially random isometries, are also scrambled as if by an approximate unitary design, regardless of $\chi$, capturing (some) elements of the volume law behavior.

To illustrate this numerically, we consider the Floquet dynamics of an $L$-site 1D kicked Ising chain~\cite{kickedising,kickedising2,kickedising3}. Expressing the Floquet evolution operator $F$ as a matrix product operator (MPO) (see SM for details), the time evolution corresponds to a (1+1)D tensor network, shown in Fig.~\ref{fig:volumelaw}(a). The tunable parameters in the model include the Ising interaction strength $J$, homogeneous transverse field $g$ and longitudinal field $h$. We focus on two representative cases for $L=14$
(i) the maximally chaotic case~\cite{selfdualkickedising}, $(J,g,h)=(\pi/4,\pi/4,0.5)$, which exhibits volume law physics,
and  (ii) the less chaotic case, $(J,g,h)=(0.7, 0.5, 0.5)$, which shows sub-volume-law physics at short times. 
For each time step $t$, we compute the bipartite von Neumann entropy $S(t)=-\mathrm{Tr}\left(\rho_A \log{\rho_A}\right)$ of the state $|\Psi(t)\rangle = F^t|0\rangle^{\otimes L}$, where $\rho_A$ is the reduced density matrix of the half-chain. (Computing $\rho_A$ is not an efficient operation, but we do so to illustrate the physics being captured). 
To define a useful TNF, we construct an contraction path that leads to configuration dependent isometries: we consider two natural ones, namely (i) contraction in the inverse time direction, and (ii) (transverse) contraction along the spatial direction. We then compare against the standard MPS-MPO compression to a bond dimension $\chi$. Results from (i), as well as from amplitude-independent MPO-MPO compression in the inverse time direction, are shown in the SM~\cite{SM}.

The evolution of $S(t)$ is depicted in the main panel of Fig.~\ref{fig:volumelaw}(b) and (c), with exact results included. In the maximally chaotic regime (parameter set $(J,g,h)=(\pi/4,\pi/4,0.5)$), which might be thought of as closest to the random circuit intuition above, the entanglement entropy grows linearly from the beginning until half-width saturation.  The required conventional MPS bond dimension to describe the full dynamics grows exponentially with time (here for a finite $L=14$ chain the maximum $\chi$ needed would be $2^{L/2} = 128$). However, volume law physics is already recovered using a TNF with $\chi=2$. The $\chi=8$ TNF accurately captures the features of the full evolution, including the maximum point in the entropy at $t=8$. Similar results are seen for the TNF defined in the inverse time-direction (see SM~\cite{SM}), although it is quantitatively less accurate.

In the less chaotic regime $(J,g,h)=(0.7, 0.5, 0.5)$,  the entanglement entropy initially grows sublinearly up until $t=7$, as shown in Fig.~\ref{fig:volumelaw}(c). The TNF for any $\chi$ captures the initial growth and more entanglement than the conventional MPS of the same bond dimension. However, at small $\chi$ the TNF entropy saturates at too small a value relative to the exact result (although the exact result also saturates below the maximum value, see SM). Nonetheless, the entanglement spectrum (insets of Fig.~\ref{fig:volumelaw}) for the TNF, even for small $\chi$, has a long tail and captures the broad features of the spectrum, rather than exhibiting the sharp cutoff of a conventional truncated MPS.

\paragraph{\it Tensor network functions can efficiently represent neural network quantum states.} As discussed TNFs only require a consistent and efficient contraction of $\Psi({\bf n})$, and there exist TNFs for which this is the case without any (amplitude dependent) isometries, but for which the efficient contraction of $\langle \Psi|\hat{O}|\Psi\rangle$ is still not possible. 
(We note that tensor networks where computing $\langle \Psi|\hat{O}|\Psi\rangle$ is efficient but $\Psi({\bf n})$ is not are used in the form of the multi-scale entanglement renormalization ansatz~\cite{vidal2007entanglement}). 
Here we describe one such application of TNFs without amplitude dependent isometries.

It has previously been observed that common tensor networks that can be efficiently exactly contracted for  $\Psi({\bf n})$ and $\langle \Psi|\hat{O}|\Psi\rangle$, i.e. MPS and tree tensor network states, can be mapped onto efficient neural networks by constructing polynomial size neural nets that emulate tensor network contraction,  establishing that exactly contractible tensor networks are a formal subclass of efficient neural quantum states~\cite{nns2022}. It has also been suggested that neural networks may not have an efficient tensor network representation, because the well known contractible tensor networks satisfy area laws, while neural networks need not do so~\cite{nns2022}.

However, if one allows for general tensor network geometries, it is straightforward to construct a TNF (without using isometries)  that represents a neural network quantum state without the restriction of an area law. We consider the explicit example of a feed forward neural network (FNN), a popular universal architecture with $k$-layers of neurons, where the outputs of each layer $\{ y^{(k-1)}_i\}$ form the inputs $ \{ x^{(k)}_i \}$  to the next layer, and the $i$-th neuron (activation) function in layer $k$ is $f^{(k)}_i(\{ x^{(k)}_{j} \}; \theta_i^{(k)})$ where $\theta_i^{(k)}$ are the neuron parameters.
Assuming that all $f$ are implemented without recursion, then the computational graph of a FNN is a directed acyclic graph. Then we observe that (i) if the FNN is efficient, the classical binary circuit implementation of the neural network function is of polynomial size in logic and copy gates, where a binary logic gate is a tensor with two binary input legs and one binary output leg, while copy takes one binary input and returns two binary outputs (nonlinearity of the activation functions enters through the copy operator) (ii) given the input product state $|{\bf n}\rangle$ (i.e. a bitstring), the application of each gate $G$ produces another product state $|{\bf n}'\rangle$ (i.e. another bitstring). Thus contracting the resulting tensor network in the appropriate order from inputs to the outputs (for example, using sparse tensor contraction, or by introducing SVDs to discover the product state structure of $|{\bf n}'\rangle$, see SM for more details~\cite{SM}) can always be done efficiently. 

The above can be seen as equating a TNF with a type of circuit computational model, in this case without a memory. 
Although the tensor network contraction is of polynomial cost, it is not very concise with respect to the number of tensors, due to representing floating point numbers in binary. In principle, we can introduce new elements to the computational model that improve the conciseness. In the Supplemental Material, we show how to use the arithmetic circuit tensor network construction of Ref.~\cite{peng2023} to compute with tensor networks with floating point entries, and the tensor network contraction is efficient if we further introduce the ability to store results from contractions of tensor network subgraphs.

\paragraph{\it Conclusions.} In summary, we have introduced tensor network functions (TNFs), a new perspective on the tensor network state, as a flexible representation for quantum many-body systems. The broadened perspective removes the conventional restrictions of tensor network computations on the form of the ansatz, and the computational need to converge the auxiliary bond dimension in an approximate contraction. On the other hand, TNFs inherit the conceptual advantages of tensor networks. Unlike generic function approximators, such as neural networks, the function architecture is directly suggested by the physics at hand and one can reuse standard tensor network algorithms. For example, in time evolution, the TNF structure can follow the evolution circuit and choice of contraction path; in variational optimization, existing TN algorithms provide a deterministic guess for the TNF tensors, which may then be further relaxed. TNFs derived from fermionic tensor networks naturally encode fermionic sign structure~\cite{fPEPS2010}, avoiding the need for more complicated constructions~\cite{RBM2017,fermionNN2022,JW2022NN,clark2023unifying}. Finally, TNFs can be further generalized into a wider class of states, where non-linearity other than the singular value decomposition is introduced into the tensor network computational graph. These possibilities open up many new avenues to investigate and to explore with  tensor networks for the future.

\paragraph{\it Acknowledgments.} This work was supported by the US National Science Foundation under grant no CHE-2102505. GKC acknowledges additional support from the Simons Investigator program and the Dreyfus Foundation under the program Machine Learning in the Chemical Sciences and Engineering.

\bibliography{TNfunction}

\onecolumngrid
\appendix
\setcounter{equation}{0}
\newpage

\renewcommand{\thesection}{S-\arabic{section}} \renewcommand{\theequation}{S%
\arabic{equation}} \setcounter{equation}{0} \renewcommand{\thefigure}{S%
\arabic{figure}} \setcounter{figure}{0}

\centerline{\textbf{Supplemental Material}}

\maketitle

\section{S-1. Monte Carlo simulations with PEPS}

\subsection{A. Variational Monte Carlo}
Here we review variational Monte Carlo  with some specific comments on its implementation with PEPS. There are standard works on this in the literature, but we follow here the presentation in Refs.~\cite{liu2017,liu2021}.

In VMC, expectation values are computed by importance sampling of configurations $\ket{{\bf n}}=|n_{1}n_{2} \cdots n_{L} \rangle$. For example, the total energy reads: 
\begin{equation}
  E_{\rm tot}= \frac{\langle \Psi |H|\Psi\rangle}{\langle \Psi | \Psi\rangle}=\frac{1}{\langle \Psi | \Psi\rangle}\sum_{{\bf n}, {\bf n}^{\prime}} \langle \Psi\ket{{\bf n}}\bra{{\bf n}} H \ket{{\bf n}^\prime}\bra{{\bf n}^\prime}\Psi\rangle=\sum_{{\bf n}}\frac{ | \langle \Psi\ket{{\bf n}}|^2}{\langle \Psi | \Psi\rangle} \sum_{{\bf n}^{\prime}} \frac{\bra{{\bf n}^\prime}\Psi\rangle}{\bra{{\bf n}}\Psi\rangle} \bra{{\bf n}} H \ket{{\bf n}^\prime} 
 = \sum_{{\bf n}}p({\bf n}) E_{\rm loc}({\bf n}) ~~,
 \label{eq:energy}
\end{equation}
where $p({\bf n})= | \langle \Psi\ket{{\bf n}}|^2 / \langle \Psi | \Psi\rangle $ is the probability of the configuration $\ket{{\bf n}}$, and $E_{\rm loc}({\bf n})$ is the local energy defined as  
\begin{equation}
E_{\rm loc}({\bf n})=\sum_{{\bf n}^{\prime}} \frac{\bra{{\bf n}^\prime}\Psi\rangle}{\bra{{\bf n}}\Psi\rangle} \bra{{\bf n}} H \ket{{\bf n}^\prime}=\sum_{\langle ij \rangle }\sum_{{\bf n}^{\prime}} \frac{\bra{{\bf n}^\prime}\Psi\rangle}{\bra{{\bf n}}\Psi\rangle} \bra{{\bf n}} H_{ij} \ket{{\bf n}^\prime} \,.
\label{eq:Es}
\end{equation}
Here $\bra{{\bf n}}\Psi\rangle$ is the amplitude of the configuration $|{\bf n}\rangle$. For simplicity we assume the Hamiltonian $H$ comprises nearest-neighbor two-site interaction terms $H_{ij}$, i.e., $H=\sum_{\langle ij \rangle} H_{ij}$. Thus in the summation of Eq.(\ref{eq:Es}) for $\ket{{\bf n}^\prime}$, we only need to consider nonzero Hamiltonian elements.
Under this assumption about the Hamiltonian, the states $\ket{{\bf n}^\prime}$ and $\ket{{\bf n}}$ can differ only at two sites.

The sampling in Eq.(\ref{eq:energy}) is performed using the standard Markov Chain Monte Carlo procedure. Starting from the current configuration $\ket{{\bf n}_{0}}$,  a trial configuration $\ket{{\bf n}_{1}}$ is 
 generated. 
 Following the  Metropolis algorithm, a trial configuration $\ket{{\bf n}_{1}}$ is then accepted as a new configuration for the Markov Chain if a randomly chosen number from the uniform distribution in the interval [0,1) is smaller than the ratio $p({\bf n}_{1})/p({\bf n}_{0})$. Otherwise,  the trial configuration $\ket{{\bf n}_{1}}$ is rejected, and another trial configuration $\ket{{\bf n}_{1}^{\prime}}$ is generated. 

The application of VMC to PEPS, either as a tensor network state or as a TNF, follows the above form. We discuss the definition of the amplitudes in the TNS and TNF in more detail in the following section. To generate configurations, we use a move which flips the spins/occupations of nearest neighbour pairs. We use a schedule of moves where we sequentially flip all allowed nearest-neighbor spin pairs in the sample subspace~\cite{liu2021}. This defines a Monte Carlo sweep. Physical quantities are measured after each Monte Carlo sweep to minimize autocorrelation, and both the energies and energy gradients are evaluated through this Monte Carlo sampling approach~\cite{liu2017,liu2021}.

We optimize the local PEPS tensors using the simple update imaginary time evolution method. This provides states close to the ground state~\cite{jiang2008}, but does not produce the variationally optimal PEPS. Subsequently, we employ gradient methods, including the stochastic gradient descent method~\cite{liu2017, liu2021} and stochastic reconfiguration optimization~\cite{SR1998, Neuscamman2012, directSam2021}, to further minimize the energy. The isometries are not directly optimized, but are obtained from the SVD of contractions of the local tensors (see next section).

\subsection{B. Definition of amplitudes in PEPS and PEPS-TNF}

\begin{figure}[htbp]
\includegraphics[width=\columnwidth]{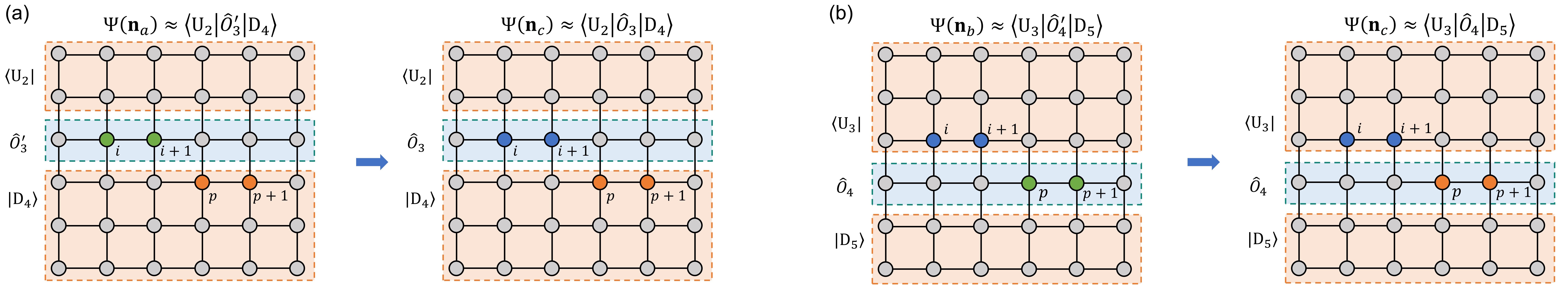}
\caption{ In a TN-VMC calculation, to reduce computational cost, the amplitude  $\Psi({\bf n}_c)$ for a configuration $\ket{{\bf n}_c}$ can be computed using different intermediates, depending on the position in the Markov chain. (a) The tensor networks for $\Psi({\bf n}_a)$ and  $\Psi({\bf n}_c)$ have two different tensors due to the differences between the configurations $\ket{{\bf n}_a}=\ket{\cdots n_{i}^\prime n_{i+1}^\prime \cdots n_p n_{p+1}\cdots}$ and $\ket{{\bf n}_c}=\ket{\cdots n_{i} n_{i+1} \cdots n_p n_{p+1}\cdots}$.  The amplitude $\Psi({\bf n}_c)$ is obtained via intermediates $\bra{U_2}$ and $\ket{D_4}$ generated in computing  $\Psi({\bf n}_a) \approx \langle {U_2} |\hat{O}_3^{\prime} |D_4 \rangle$, and we compute $\Psi({\bf n}_c) \approx \langle {U_2} |\hat{O}_3 |D_4 \rangle$. (b) The tensor networks for $\Psi({\bf n}_b)$ and  $\Psi({\bf n}_c)$ have two different tensors. $\Psi({\bf n}_c)$ is obtained via intermediates  $\bra{U_3}$ and $\ket{D_5}$ generated in computing  $\Psi({\bf n}_b)\approx\langle {U_3} |\hat{O}_4^{\prime} |D_5 \rangle$, and it has
$\Psi({\bf n}_c)\approx\langle {U_3} |\hat{O}_4 D_5 |\rangle$. The intermediates $\bra{U_2}$, $\ket{D_4}$, $\bra{U_3}$ and $\ket{D_5}$ are obtained by the standard boundary-MPS contraction method (see next section). }
\label{fig:VMCPEPS} 
\end{figure}

The key difference between the standard view of a PEPS and the TNF view, is that in the standard view, contracting the PEPS is necessarily approximate for a general graph, i.e. it must have some error for any reasonable computation. This means that one can choose to approximately contract the PEPS in different ways (i.e. insert isometries in different ways), so long as the approximation error is tolerable. Consequently, in the standard use of PEPS with VMC, one is free to choose different contraction schemes depending on where one is in the Markov chain in order to reuse as many intermediates possible and reduce the computational cost, so long as the error is comparable for each contraction scheme. On the other hand, in a TNF, the position of the isometries is part of the definition of the exact TNF, so one is not free to change them, otherwise one is changing the type of TNF one is handling.

As an example of reusing intermediates in standard TN-VMC, we show in Fig.~\ref{fig:VMCPEPS}(a) the evaluation of the amplitude $\Psi(\mathbf{n}_c)$ where the configuration $\ket{{\bf n}_c}$ is obtained from a Monte Carlo move from $\ket{{\bf n}_a}$. 
$\ket{{\bf n}_a}=\ket{\cdots n_{i}^\prime n_{i+1}^\prime \cdots n_p n_{p+1}\cdots}$ and differs from 
$\ket{{\bf n}_c}=\ket{\cdots n_{i} n_{i+1} \cdots n_p n_{p+1}\cdots}$ only at the sites $i$ and $i+1$, so the computation of the amplitude $\Psi({\bf n}_c)$ can reuse much of the computation involved in computing $\Psi({\bf n}_a)$. We can arrange for this reuse by building the intermediates involved by approximately contracting tensors from the top and bottom respectively. Fig.~\ref{fig:VMCPEPS}(b) shows another example of this reuse in action. Here we consider the same $\ket{{\bf n}_c}$, but this time it is reached by a Monte Carlo move from a prior configuration $\ket{{\bf n}_b}$, which differs at sites $p$ and $p+1$ from $\ket{{\bf n}_c}$. Consequently, we compute $\Psi({\bf n}_c)$ using different intermediates obtained by contracting the tensors in a different order. The
use of different intermediates in Fig.~\ref{fig:VMCPEPS}(a) and Fig.~\ref{fig:VMCPEPS}(b) mean that the locations of the isometries involved in defining the same amplitude $\Psi({\bf n}_c)$ is different in the two cases.

\begin{figure}[htbp]
\includegraphics[width=\columnwidth]{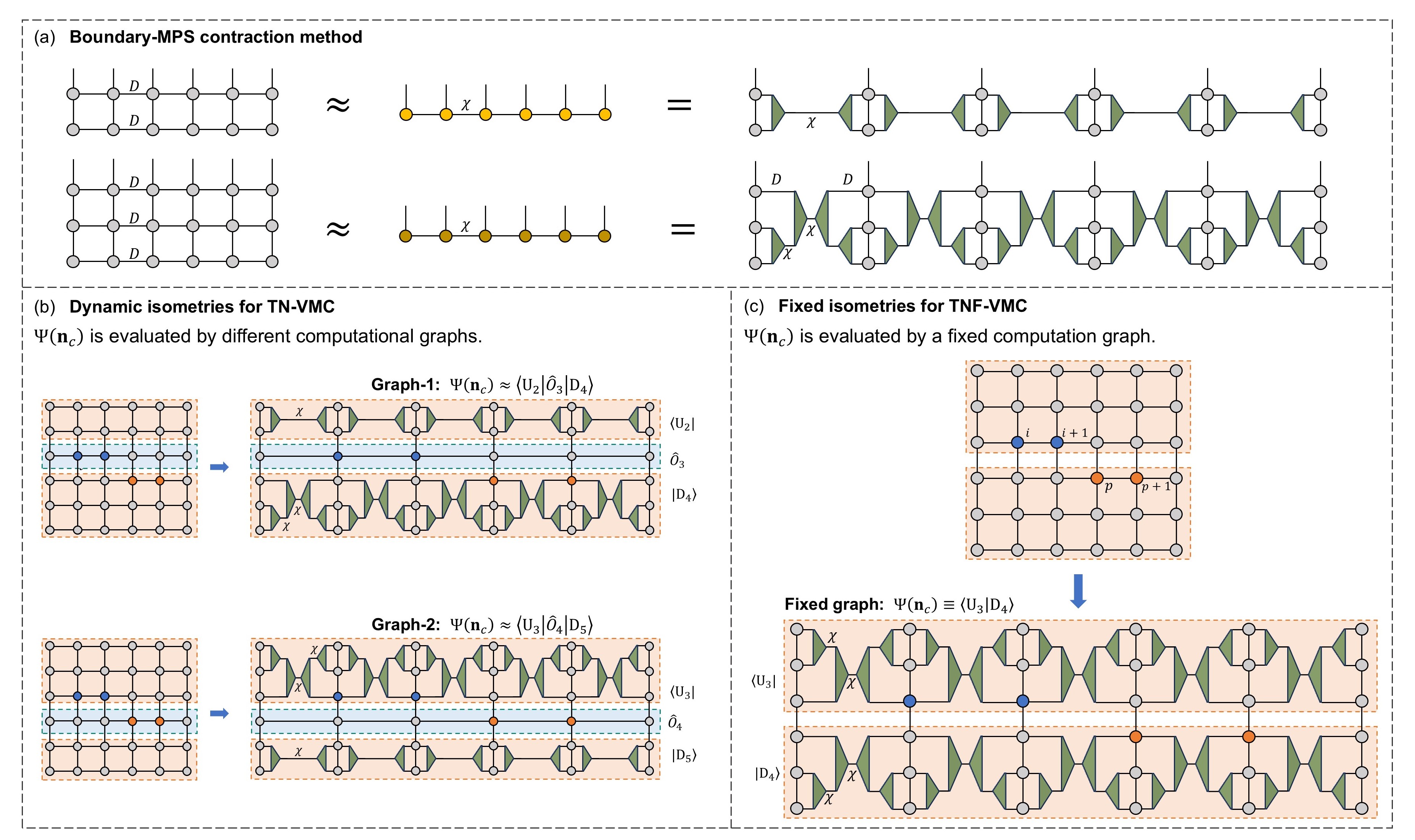}
\caption{ (a) The boundary-MPS method is used, and a two-row tensor network (contracting from bottom up) with bond dimension $D$ is compressed into an MPS with bond dimension $\chi$, which can be realized by inserting isometries obtained from SVD. Similarly, a three-row tensor network (contracting from bottom up) is first compressed into a two-row tensor network and then  further compressed into an MPS, which can be represented as a tensor network contraction by inserting two layers of isometries obtained by SVDs.  Note here  ``$\approx$'' indicates approximate tensor network contraction controlled by $\chi$, and it would be rigorously  ``$=$'' in the limit $\chi\rightarrow \infty$.  (b) Given the PEPS, in a TN-VMC simulation, for a given configuration $\ket{{\bf n}_c}$, the amplitude is evaluated by different computational graphs, depending on its prior configurations such as $\ket{{\bf n}_a}$ and $\ket{{\bf n}_b}$. (c) Given the PEPS, in a TNF-VMC simulation,  for a  configuration $\ket{{\bf n}_c}$, its amplitude is always evaluated by a fixed computational graph. }
\label{fig:PEPSgraph} 
\end{figure}

The specific choice of isometries used in the calculations for Fig. 2 and Fig. 3 in the main text (associated with PEPS on a square lattice)  are shown in Fig.~\ref{fig:PEPSgraph}. These isometries and their locations follow the SVD compression steps of the standard boundary-MPS contraction method. On the left hand side, we show the isometries used in TN-VMC where different isometries are used  at different points in the Markov chain as discussed above (corresponding to the blue site spins being flipped and the orange site spins being flipped). On the right hand side we show the isometries used to define the TNF. Here the positions are kept fixed for every amplitude, and the values of the isometries are obtained in the boundary-MPS contraction and compression steps from the top and the bottom up to the middle layer of the tensor network, regardless of which Markov chain step/configuration we are considering.

\section{S-2. 1D kicked Ising model dynamics}
\subsection{A. Definition of the model}
\begin{figure}[htb]
\centering

\subfloat[]{%
  \includegraphics[width=0.32\columnwidth]{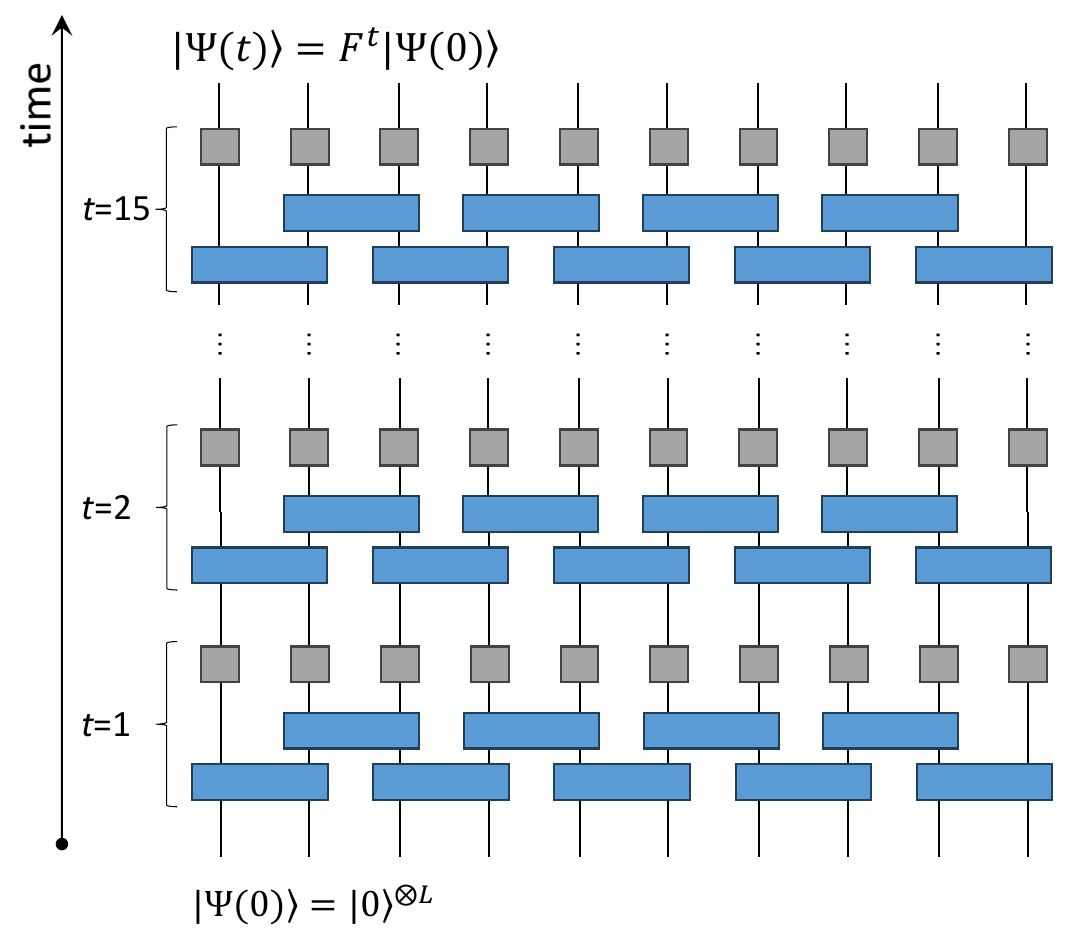}%
}\hspace*{\fill}%
\subfloat[]{%
  \includegraphics[width=0.32\columnwidth]{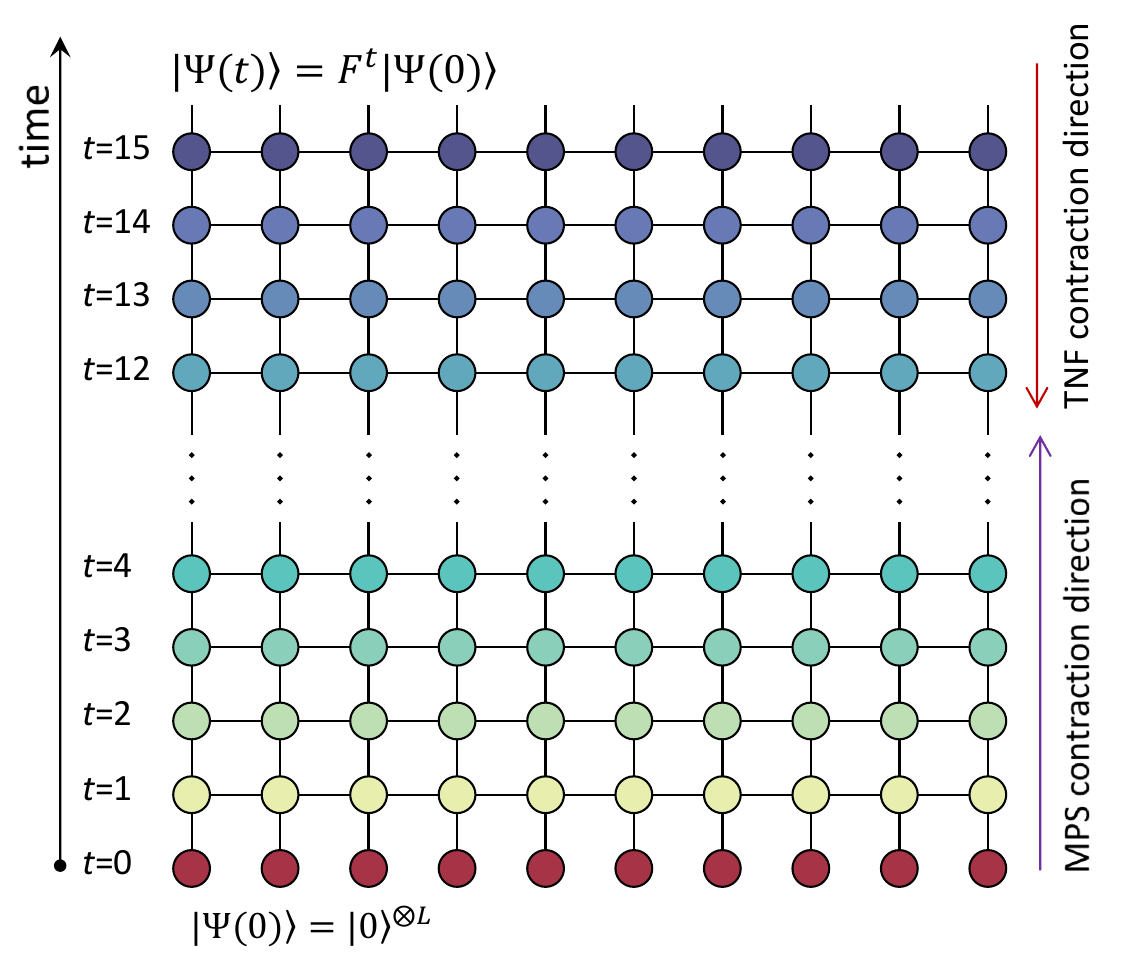}%
}\hspace*{\fill}%
\subfloat[]{%
  \includegraphics[width=0.32\columnwidth]{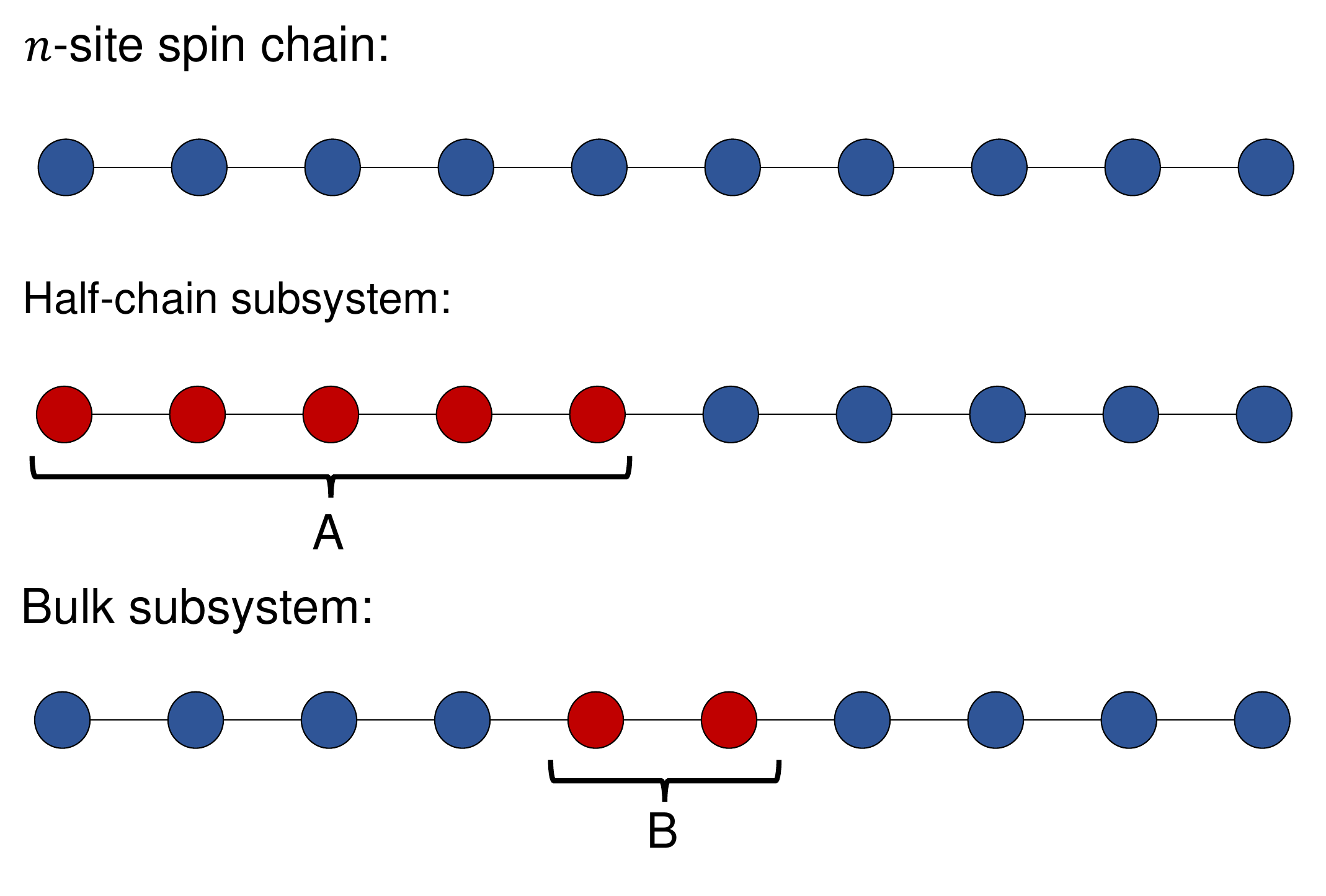}%
}\hspace*{\fill}%

\caption{\label{fig:2dtn}
(a) The quantum circuit representation of the kicked Ising Floquet dynamics. (b) The $(1+1)$D tensor network for Floquet dynamics, with contraction order either along the inverse-time direction for tensor network function or time-evolution direction for standard MPS-MPO compression. (c) The 1D spin chain and different subsystems to compute the entanglement entropy.}
\end{figure}

Here we give the detailed definition of the 1D kicked Ising model and its corresponding unitary Floquet dynamics~\cite{kickedising,kickedising2,kickedising3} discussed in the main article. Starting from a product state $|0\rangle^{\otimes L}$ in an $L$-site spin chain,  we consider the time evolution of the kicked Ising Hamiltonian $H(t)=H_{\mathrm{I}} + H_{\mathrm{K}}\sum_{m=-\infty}^{\infty}\delta(t-m)$, where $\delta(t)$ is the Dirac $\delta$ function, with
 \begin{align}
     H_{\mathrm{I}} &= -J\sum_{j=1}^{L-1} \sigma^z_{j}\sigma^z_{j+1} - h\sum_{j=1}^{L} \sigma^z_j \quad, \\
     H_{\mathrm{K}} &= -g\sum_{j=1}^L \sigma^x_j \quad.
 \end{align}
 Setting the time interval between kicks to $1$, the unitary Floquet operator generated by the above Hamiltonian reads as (set $\hbar = 1$):
 \begin{align}
     F_{\mathrm{KI}} &= e^{-i H_{\mathrm{K}}}e^{-i H_{\mathrm{I}}}\\
     & =\prod_{j} e^{i g \sigma^x_j} \prod_{j}e^{i h \sigma^z_j} \prod_{j\ even}e^{i J\sigma^z_{j}\sigma^z_{j+1} } \prod_{j\ odd} e^{i J\sigma^z_{j} \sigma^z_{j+1}} \quad, \label{eq:kickedising}
\end{align}
which admits a straightforward quantum circuit representation with only single-qubit and two-qubit gates, as shown in Fig.~\ref{fig:2dtn}(a). By combining the unitary gates within each Floquet time step into an MPO (where we use singular value decomposition to split the two-qubit gates into local MPOs and contract tensors on each site), we can represent the Floquet evolution as a $(1+1)$D tensor network as illustrated in Fig.~\ref{fig:2dtn}(b).

\subsection{B. Numerical results on the 1D kicked Ising model dynamics}
\subsubsection{1. Choice of model parameters}
 To study the quantum (pure) state entanglement in the kicked Ising model dynamics, for each Floquet time step $t$, we compute the half-chain bipartite von Neumann entropy $S(t)=-\mathrm{Tr}\left(\rho_A \log{\rho_A}\right)$ of the state $|\Psi(t)\rangle = F^t|0\rangle^{\otimes L}$ {(where $A$ indicates the subsystem of the half-chain (the second row in Fig.~\ref{fig:2dtn}(c)))} and record its value along the Floquet evolution. The reduced density matrix is obtained as the partial trace of the whole density matrix $\rho=\sum_{ij}\langle \vec{n}_i|\Psi\rangle\langle\Psi|\vec{n}_j\rangle|\vec{n}_i\rangle\langle \vec{n}_j|$, which is constructed by enumerating all amplitudes $\langle \vec{n}_i|\Psi\rangle$ (assuming normalization) in the $\sigma^z$ basis for the system sizes considered in this work. We choose two representative sets of parameters $(J,g,h)=(\pi/4, \pi/4, 0.5)$ and $(J,g,h)=(0.7, 0.5, 0.5)$. Both parameter sets correspond to non-integrable dynamics but have different entanglement growth speeds. For $(J,g,h)=(\pi/4, \pi/4, 0.5)$, the kicked Ising Floquet dynamics is maximally chaotic where the half-chain entanglement entropy grows at a maximal speed and linearly in time~\cite{selfdualkickedising}. For $(J,g,h)=(0.7, 0.5, 0.5)$, the dynamics is less chaotic and has a slower half-chain entanglement growth speed. Results of the exact entanglement dynamics for both parameter sets are shown in Fig.~\ref{fig:volumelaw} in the main article as well as in Fig.~\ref{fig:inverse}. Here we point out that the choice of the parameter values for the less chaotic case is not fine-tuned, but rather an arbitrary choice that is away from the maximally chaotic regime. We expect similar conclusions (about tensor network function) to hold for other general choices of parameters in the 1D kicked Ising dynamics away from the maximally chaotic regime.

\subsubsection{2. Spatial entanglement structure of the evolved states}
\begin{figure}[htb]
\centering

\subfloat[]{%
  \includegraphics[width=0.48\columnwidth]{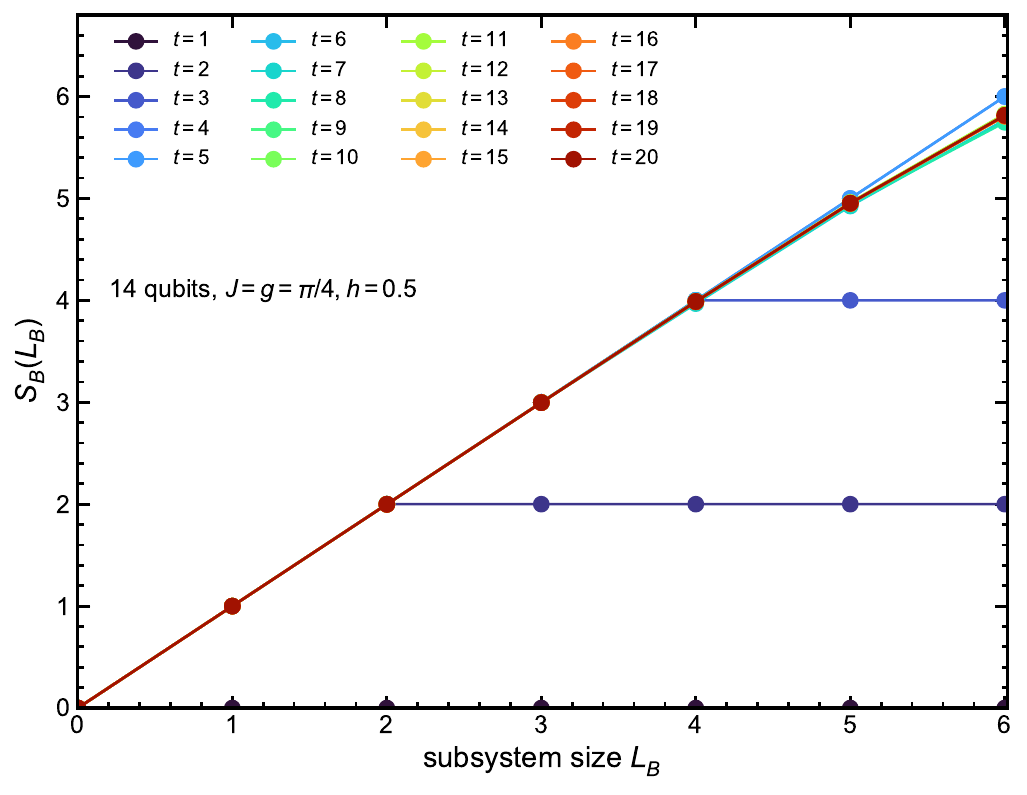}%
}\hspace*{\fill}%
\subfloat[]{%
  \includegraphics[width=0.48\columnwidth]{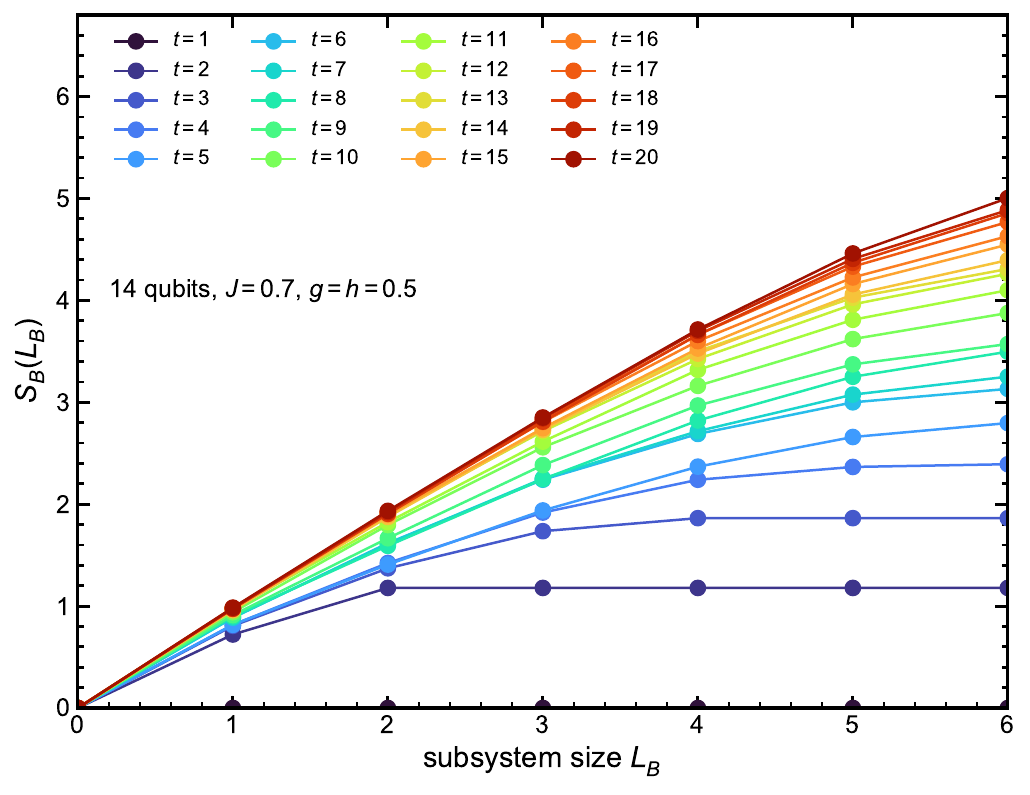}
}\hspace*{\fill}%

\caption{\label{fig:volumelaw_si} Bulk subsystem entanglement entropy scaling with respect to the subsystem size along the kicked Ising model dynamics, obtained from exact state vector simulation for $20$ Floquet steps in a $L=14$ spin chain. (a) Maximally chaotic regime ($J=g=\pi/4, h=0.5$), where evolved states after $t=3$ already display volume-law behavior. (b) Less chaotic regime ($J=0.7, g=h=0.5$), where many of the evolved states in early steps display area-law entanglement entropy behavior.} 
\end{figure}

 To demonstrate the spatial entanglement structure of the evolved states along the kicked Ising dynamics, we consider a bulk subsystem $B$ in the middle of a spin chain (the third row in Fig.\ref{fig:2dtn}(c)) of size $L=14$, and compute the entanglement entropy $S_B(L_B)$ of its reduced density matrix $\rho_B$ as a function of the subsystem size $L_B$ for Floquet time up to $t=20$. For the maximally chaotic case $(J,g,h)=(\pi/4, \pi/4, 0.5)$, after a very short evolution time, the state starts to display volume-law physics where the bulk subsystem entanglement entropy grows linearly with the subsystem size $L_B$, as shown in Fig.~\ref{fig:volumelaw_si}(a). For the less chaotic case $(J,g,h)=(0.7, 0.5, 0.5)$, many early-evolved states obey an entanglement entropy's area law where the bulk subsystem entropy saturates as the subsystem size increases, as shown in Fig.~\ref{fig:volumelaw_si}(b).

 In the maximally chaotic regime, an accurate conventional MPS simulation requires a bond dimension $\chi$ that grows exponentially with the system size, even for a short-time simulation. In the less chaotic case, however, a conventional MPS with a finite constant bond dimension $\chi$ can still describe the evolved state in the early-time evolution regime.

\subsubsection{3. Longer time entanglement dynamics for the less chaotic regime}
\begin{figure}[htb]
\centering

\subfloat[]{%
  \includegraphics[width=0.9\columnwidth]{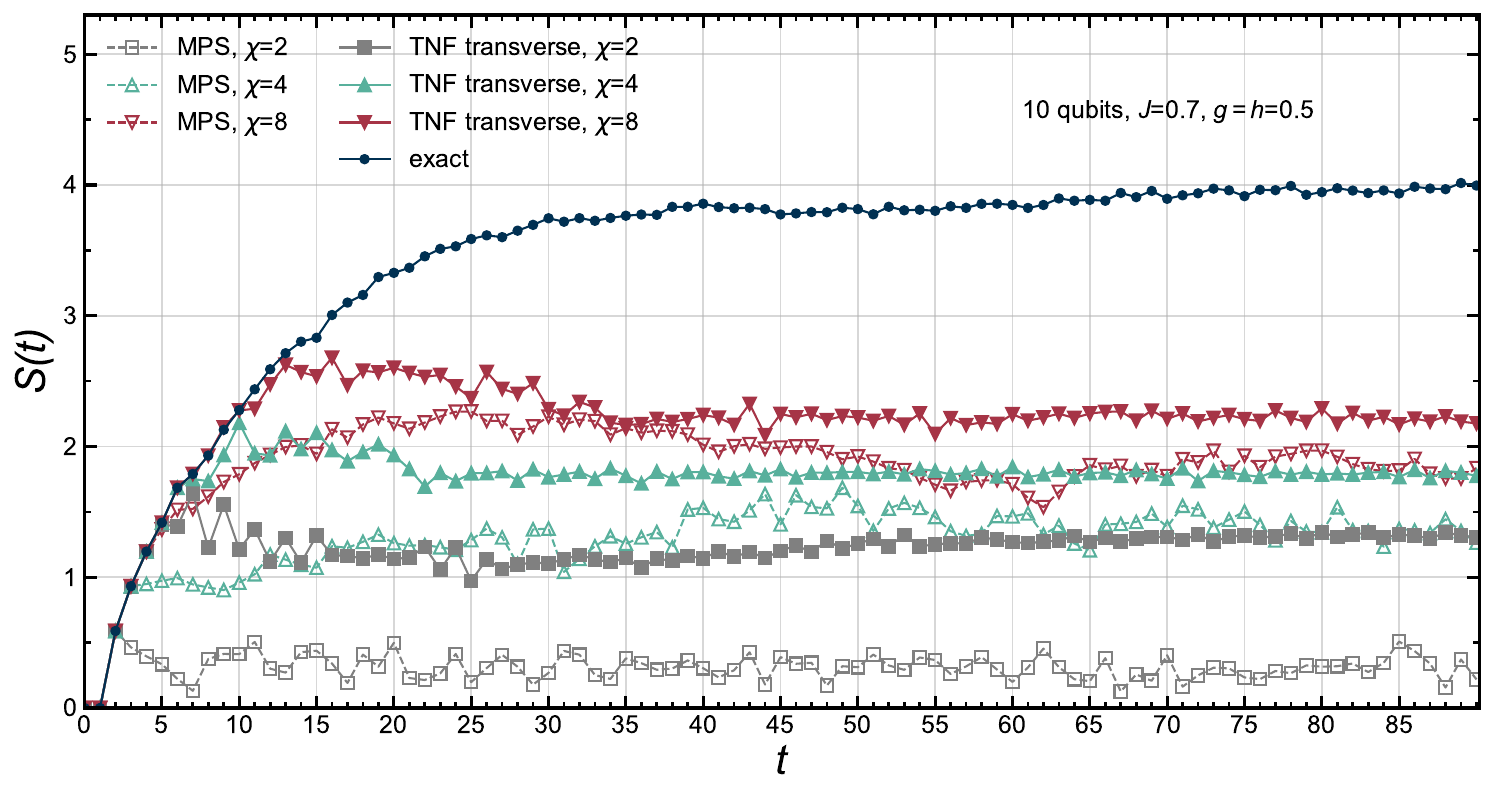}%
}\hspace*{\fill} \\ 
\subfloat[]{%
  \includegraphics[width=0.9\columnwidth]{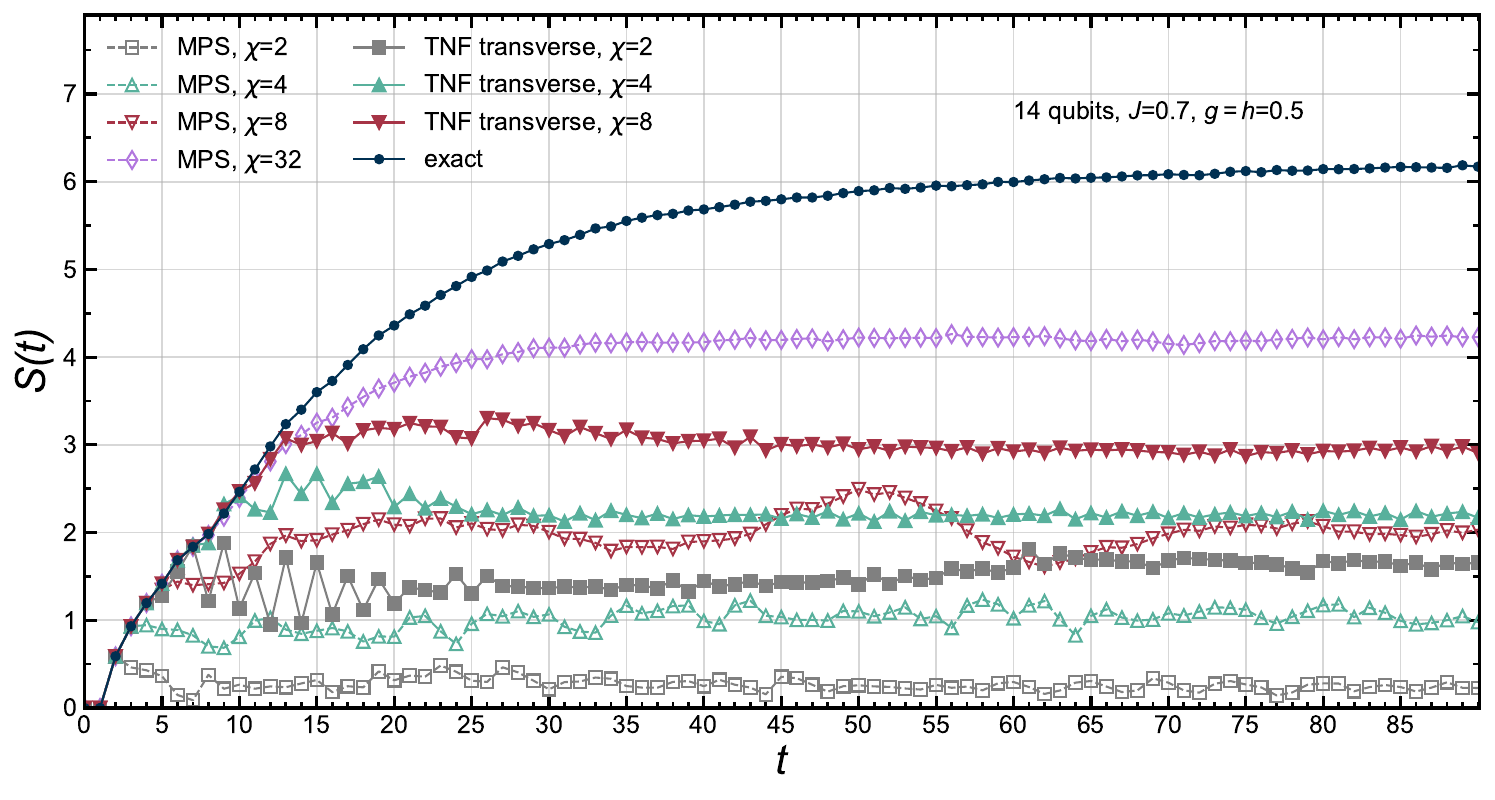}
}\hspace*{\fill}%

\caption{\label{fig:longertime} Entanglement dynamics of the kicked Ising chain in the less chaotic regime for Floquet time up to $t=90$ for (a) $L=10$ chain and (b) $L=14$ chain. Transverse contraction is used for the TN function, and conventional MPS-MPO contraction along the time direction is used for the MPS results.} 
\end{figure}

In the Fig.~\ref{fig:volumelaw}(c) in the main article where the dynamics is less chaotic at $(J,g,h)=(0.7, 0.5, 0.5)$, the half-chain entanglement entropy of the tensor network function appears to saturate at Floquet time $t=15$. To show that the entanglement entropy grows very slowly at long times, we present more calculation results up to Floquet time $t=90$ for $L=10$ and $L=14$ chains, as shown in Fig.~\ref{fig:longertime}.

\subsubsection{4. Half-chain entanglement spectrum}
\begin{figure}[htb]
\centering
   
\subfloat[]{%
  \includegraphics[width=0.48\columnwidth]{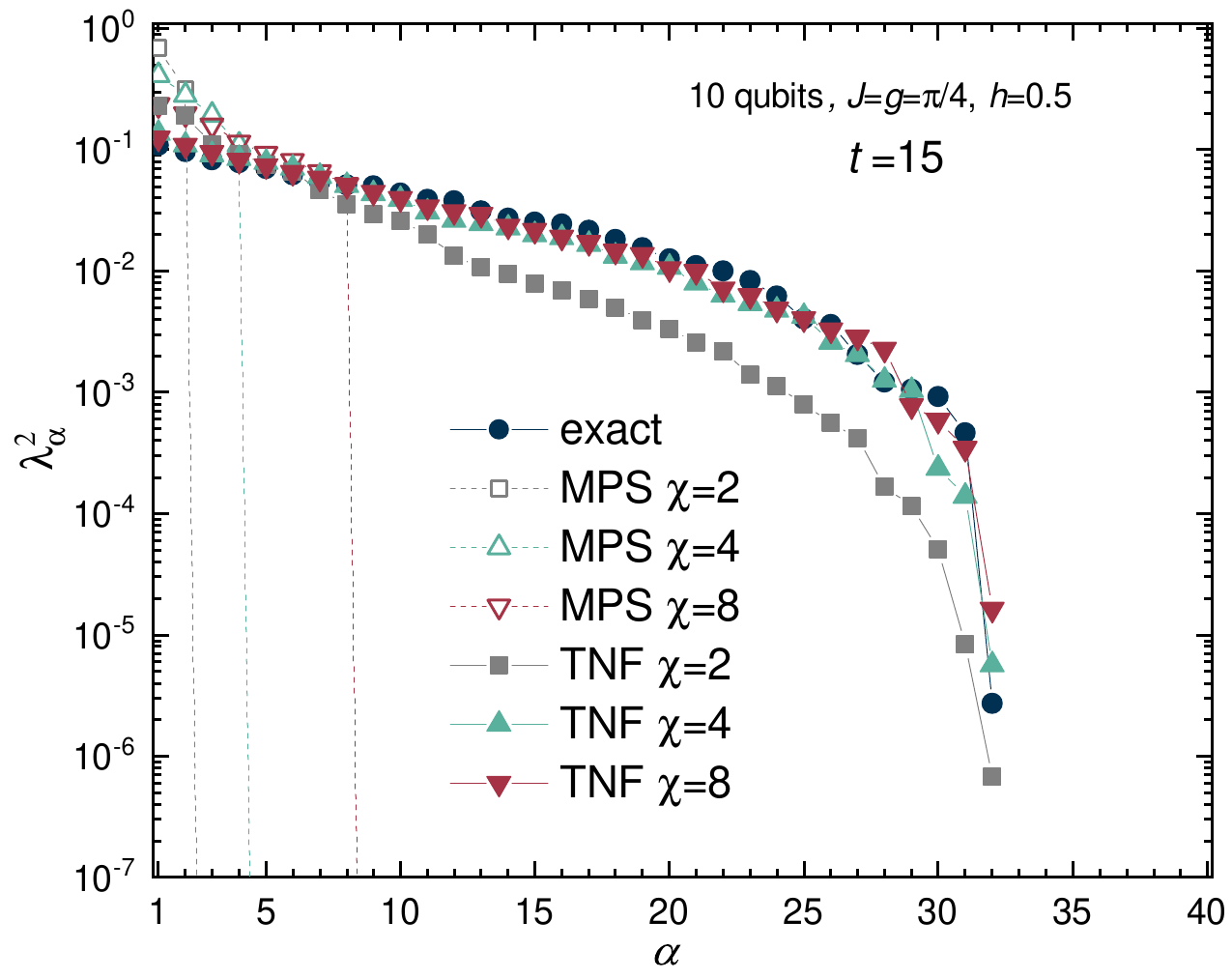}%
}\hspace*{\fill}%
\subfloat[]{%
  \includegraphics[width=0.48\columnwidth]{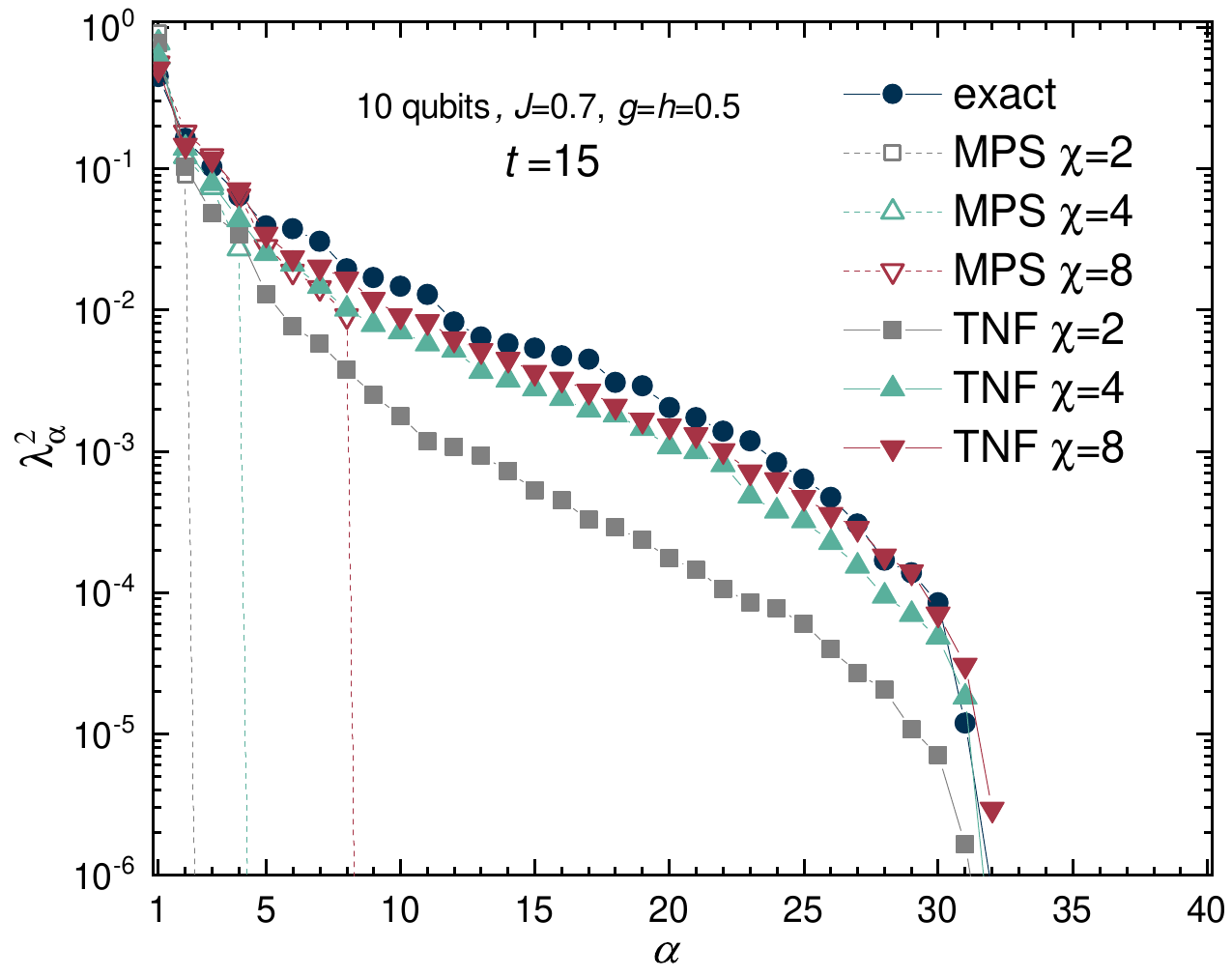}%
}\hspace*{\fill}   \\
\subfloat[]{%
  \includegraphics[width=0.48\columnwidth]{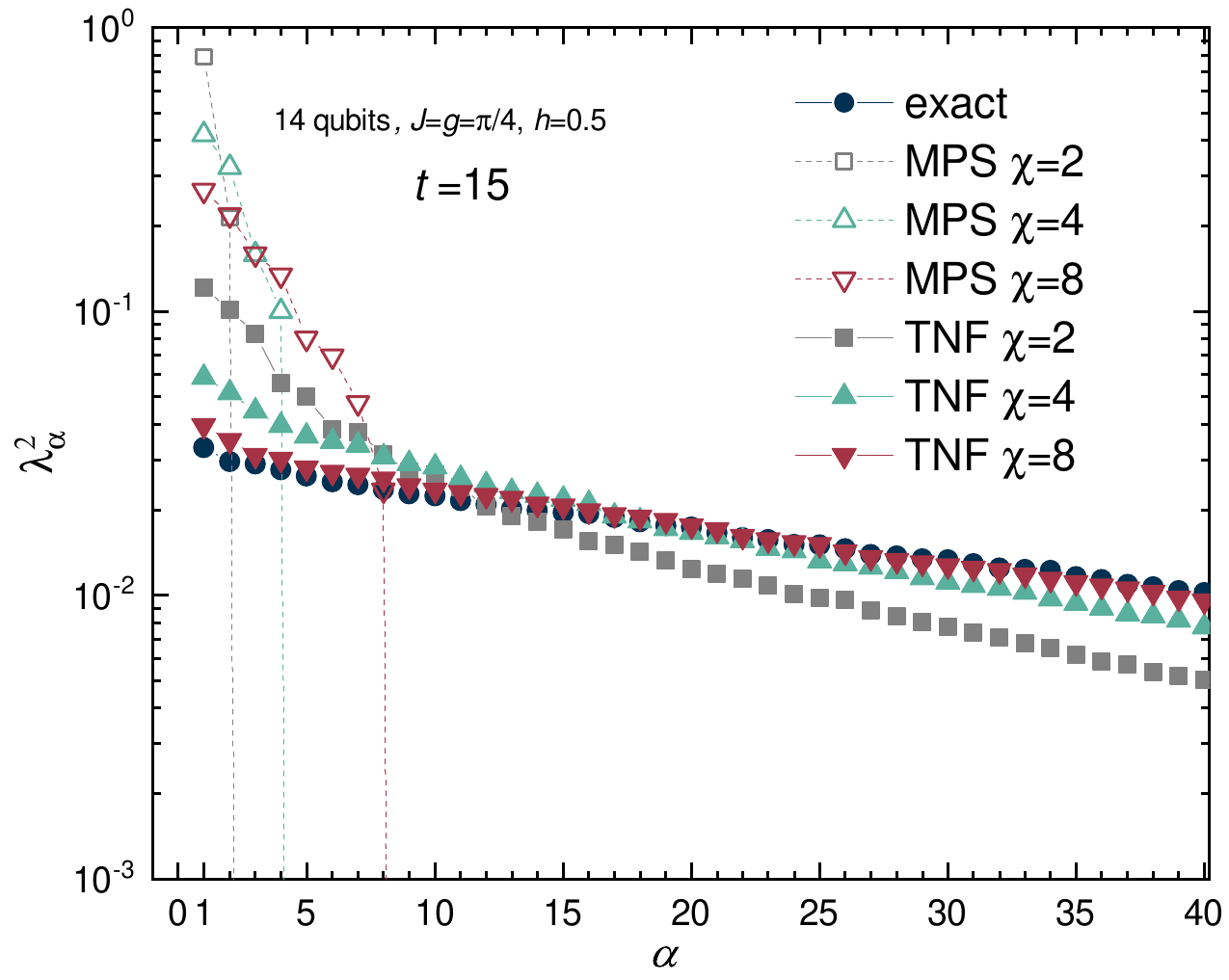}%
}\hspace*{\fill}%
\subfloat[]{%
  \includegraphics[width=0.48\columnwidth]{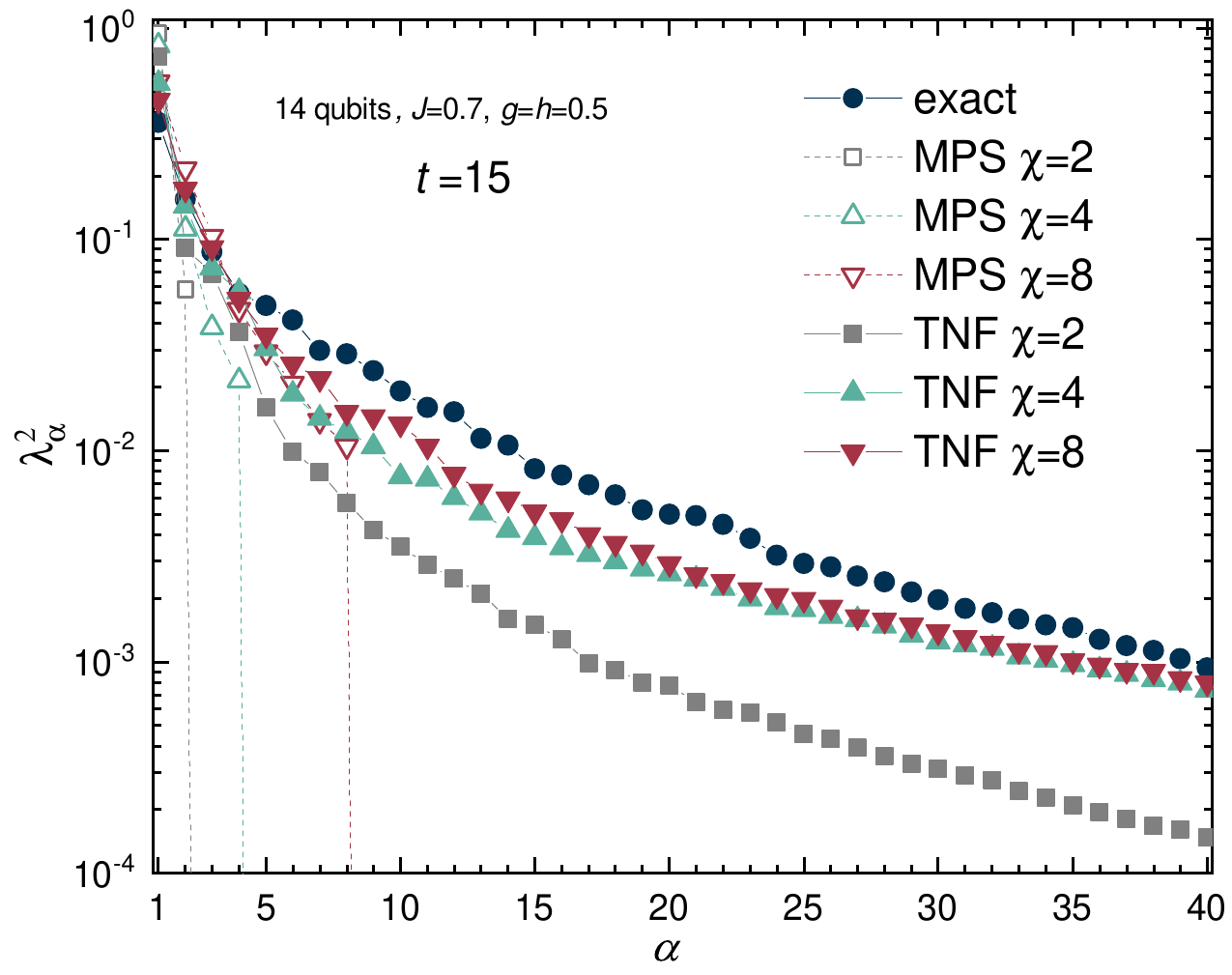}%
}\hspace*{\fill}%

\caption{\label{fig:espectrum} Half-chain entanglement spectrum for $L = 10$ (a)-(b) and $L = 14$ (c)-(d) kicked Ising chains at Floquet step $t=15$. $\lambda^2_\alpha$ is the $\alpha$-th largest eigenvalue of the half-chain reduced density matrix, and 40 largest eigenvalues are shown here. (a) and (c): Maximally chaotic regime ($J=g=\pi/4,\ h=0.5$); (b) and (d): Less chaotic regime ($J=0.7,g=0.5,h=0.5$). Transverse contraction is used for the TN function, and conventional MPS-MPO contraction along the time direction is used for the MPS results.} 
\end{figure}

To further analyze the entanglement structure captured by tensor network function, we compute the entanglement spectrum of the half-chain reduced density matrix at Floquet step $t=15$ for the $L=10$ and $L=14$ chains, comparing the results of the conventional MPS and tensor network function. More specifically, we evolve a product state $|0\rangle^{\otimes L}$ under the kicked Ising dynamics (Eq.~\ref{eq:kickedising}) for $t=15$ steps, then compute the eigenvalues $\lambda^2_\alpha$ of the half-chain reduced density matrix $\rho_A = \sum_\alpha \lambda^2_\alpha |\alpha\rangle\langle \alpha|$ of the final state using the conventional MPS and tensor network function. Exact state vector simulation is also performed to provide benchmark. Results are shown in Fig.~\ref{fig:espectrum}. The tensor network function is contracted transversely along the spatial direction. Figs.~\ref{fig:espectrum}(c) and (d) are clearer versions of the insets of Figs. 4(b) and (c) in the main article. As shown, the tensor network function with small isometry bond dimension $\chi$ can already capture the broad features of the spectrum, while the conventional MPS with bond dimension $\chi$ exhibits a sharp cutoff at the $\chi$-th largest eigenvalue.

\subsubsection{5. Entanglement dynamics from inverse-time contraction}

The tensor network function formalism allows for very flexible ways to define the amplitude from the tensor network graph, depending on how one inserts the isometries. Each choice of isometry insertion  yields a different tensor network function. 
In the numerical data shown in the main article, we employed a transverse contraction along the spatial direction for the calculation of amplitudes. Here we perform the same half-chain entanglement dynamics calculation using a tensor network function corresponding to contraction along the inverse-time direction, as illustrated in Fig.~\ref{fig:2dtn}(b). 

\begin{figure}[htb]
\centering

\subfloat[]{%
  \includegraphics[width=0.24\columnwidth]{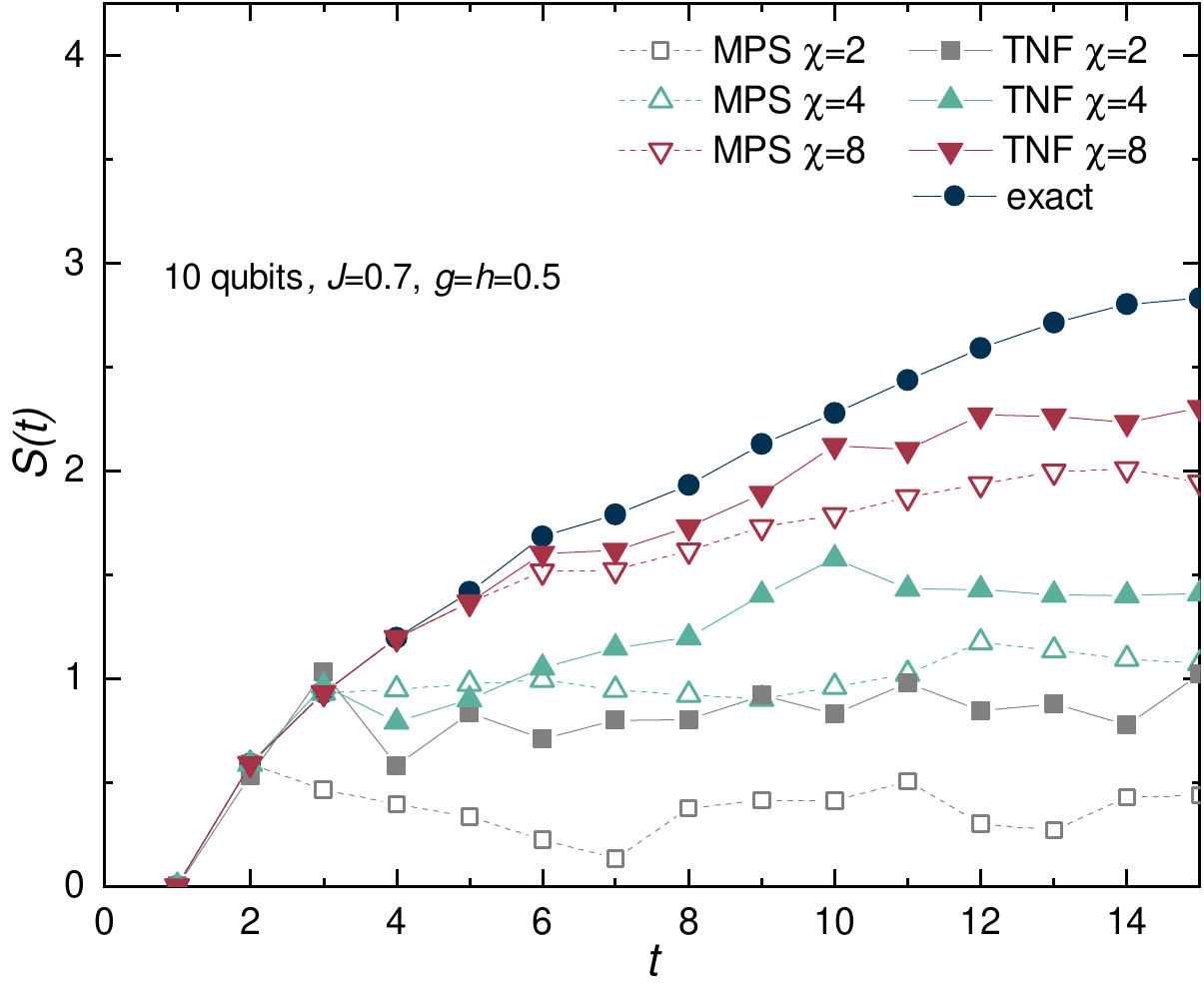}%
}\hspace*{\fill}%
\subfloat[]{%
  \includegraphics[width=0.24\columnwidth]{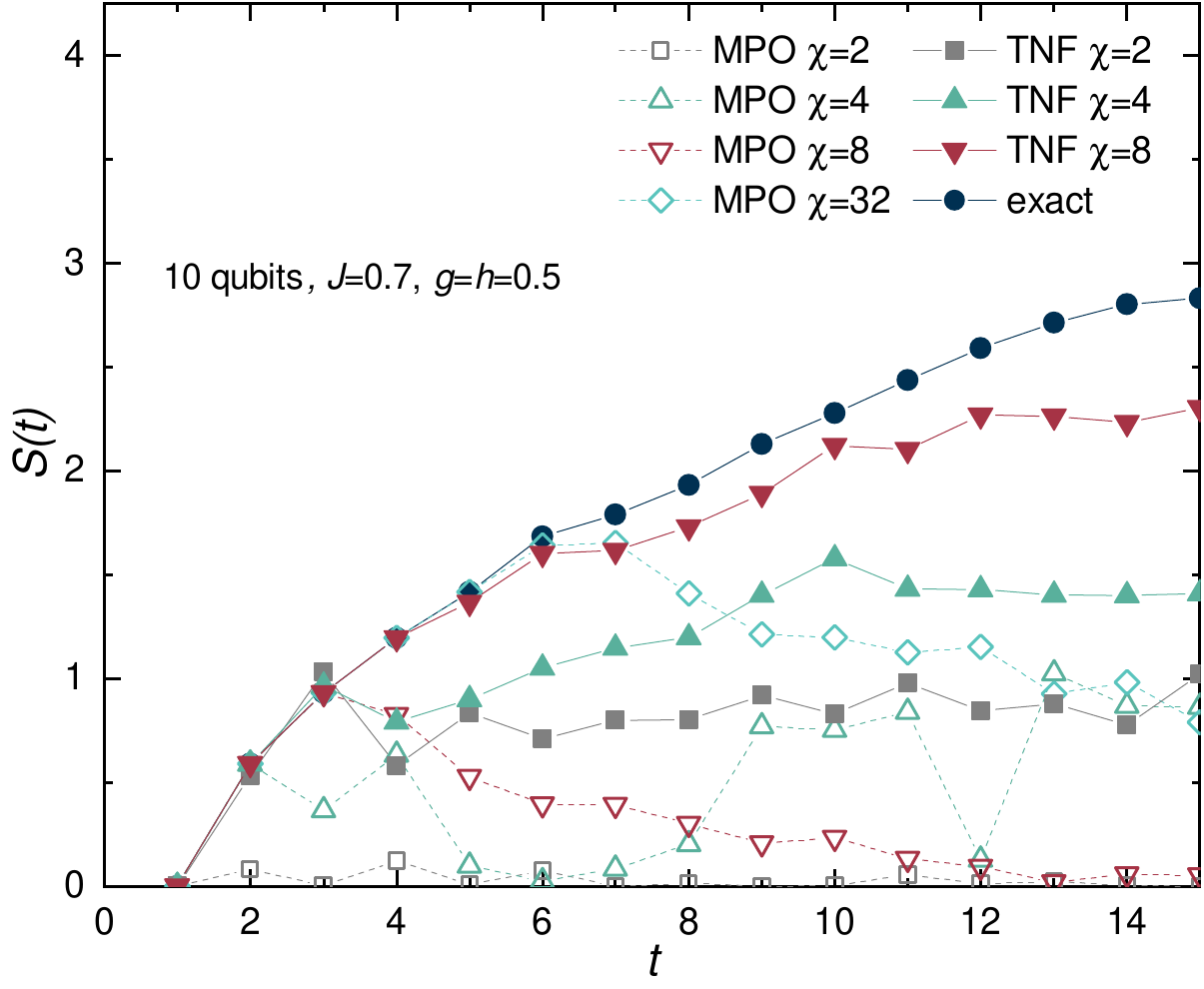}%
}\hspace*{\fill}   
\subfloat[]{%
  \includegraphics[width=0.24\columnwidth]{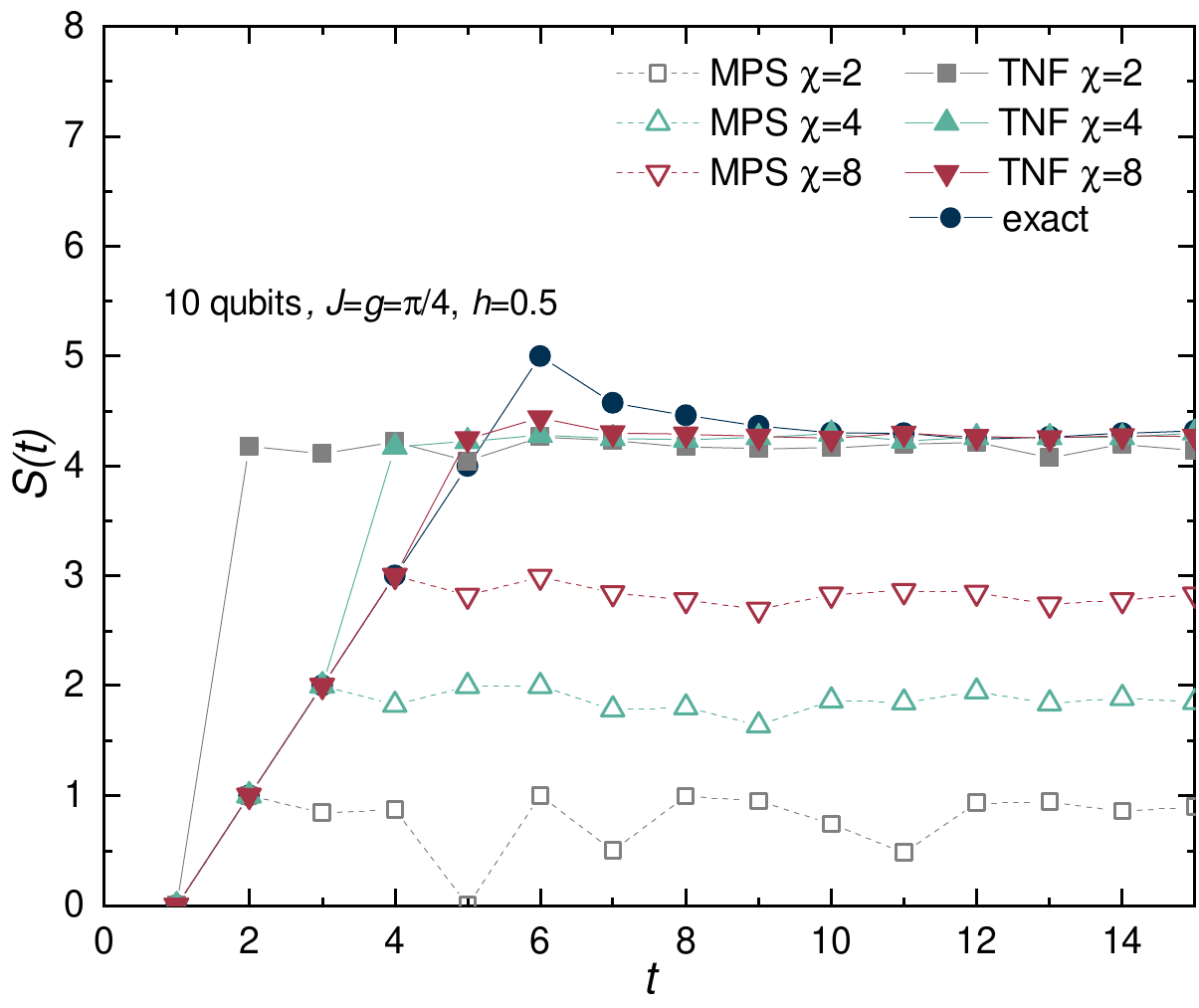}
}\hspace*{\fill}%
\subfloat[]{%
  \includegraphics[width=0.24\columnwidth]{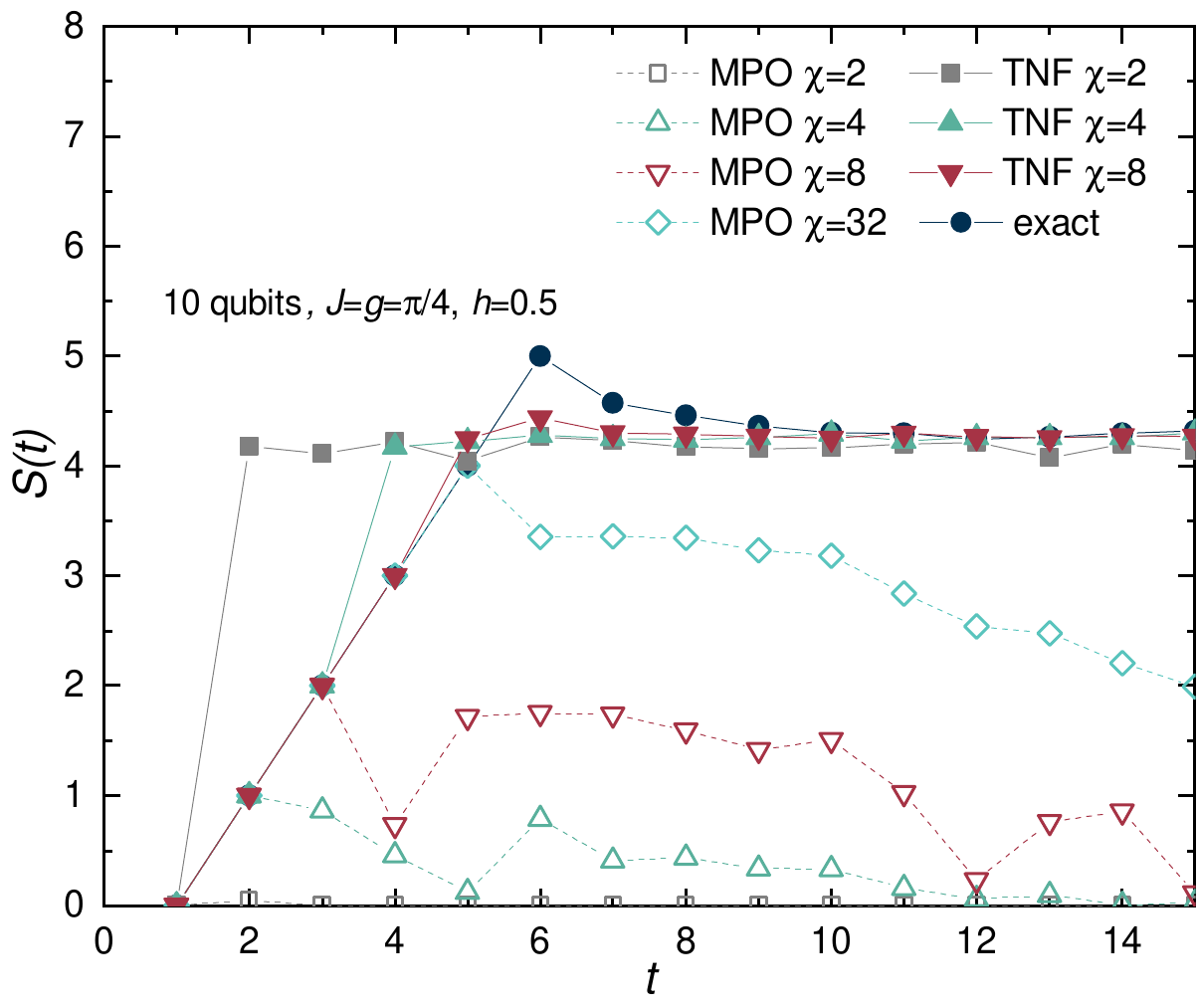}
}\hspace*{\fill}   \\

\subfloat[]{%
  \includegraphics[width=0.24\columnwidth]{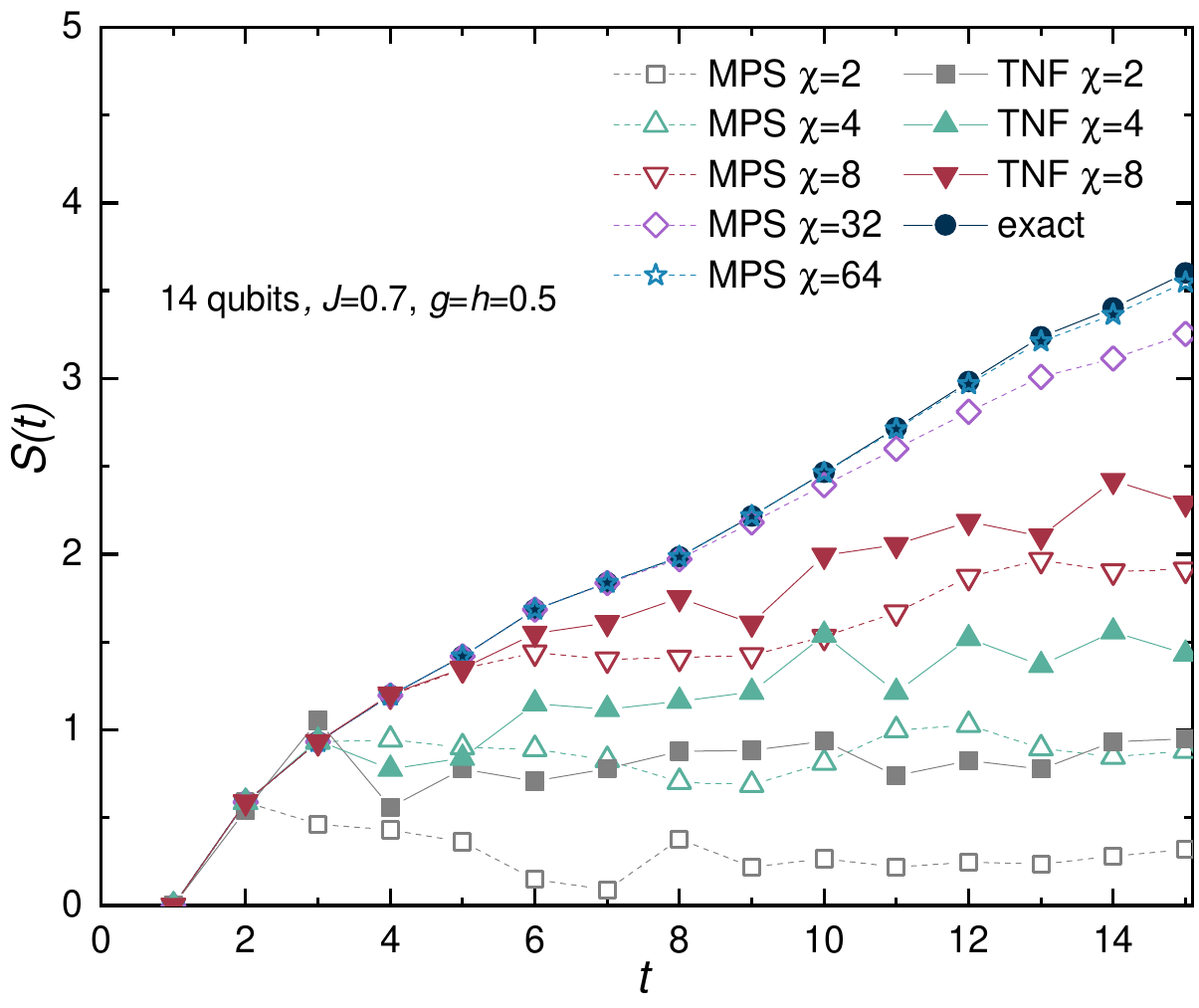}%
}\hspace*{\fill}%
\subfloat[]{%
  \includegraphics[width=0.24\columnwidth]{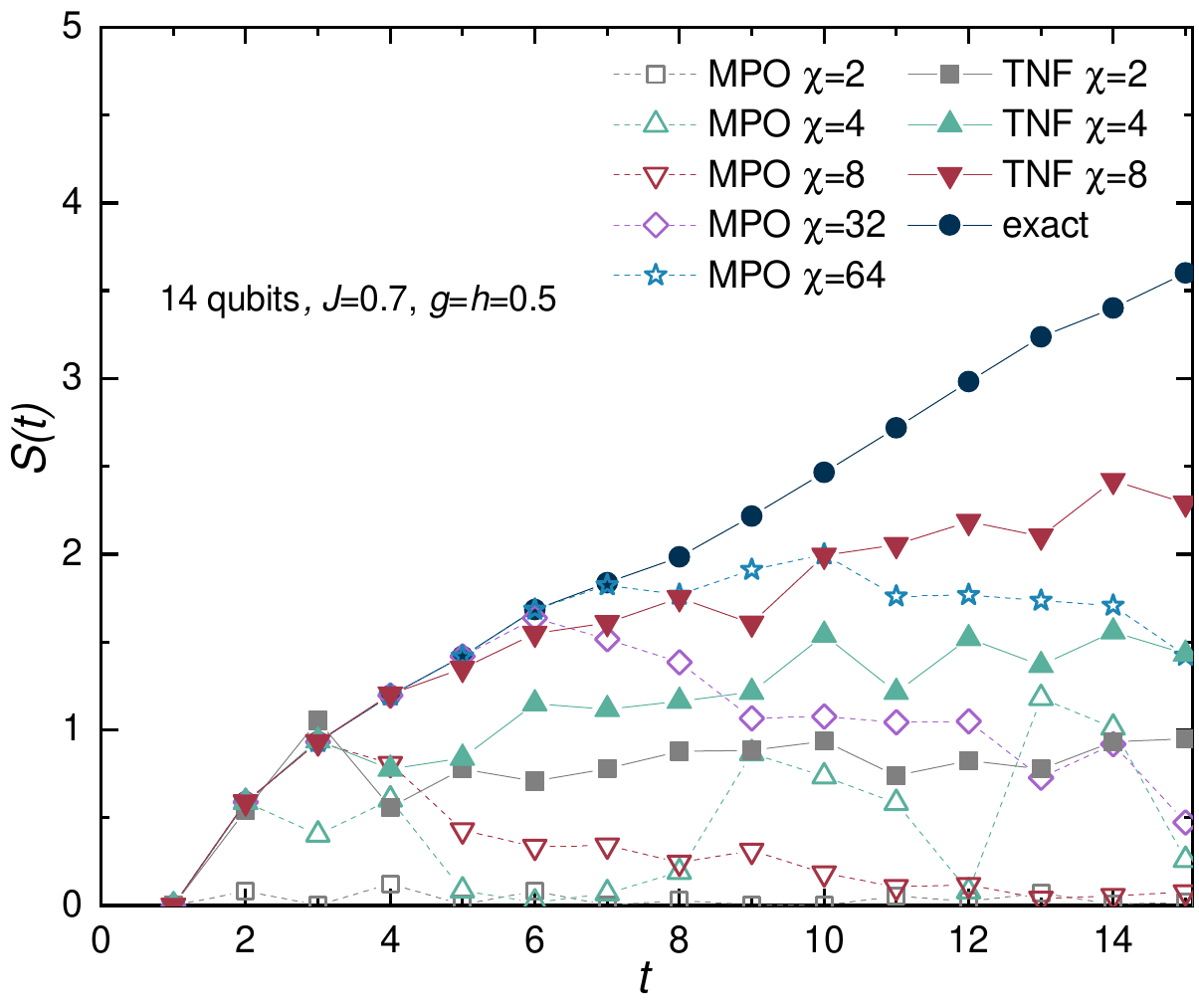}
}\hspace*{\fill}   
\subfloat[]{%
  \includegraphics[width=0.24\columnwidth]{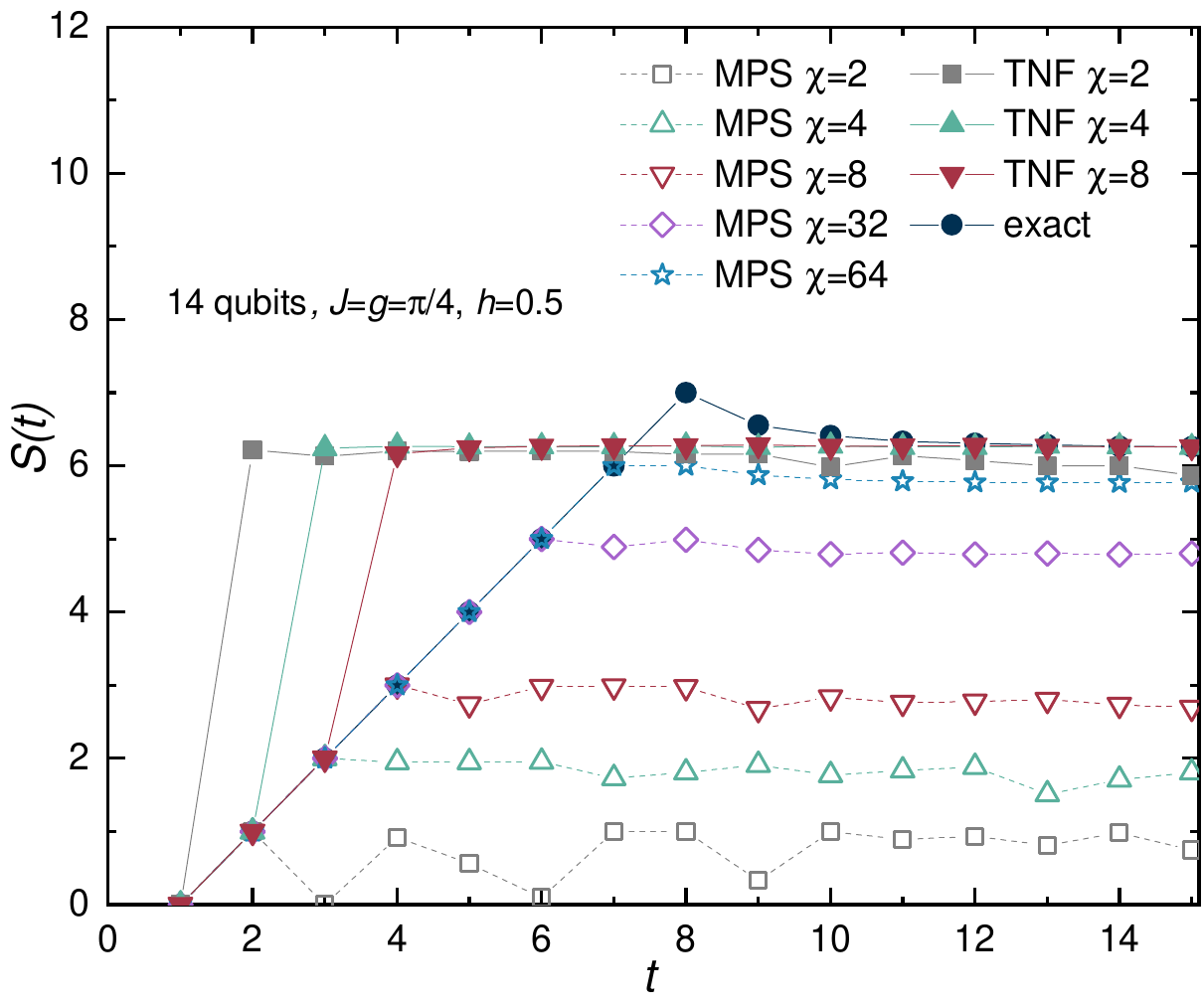}
}\hspace*{\fill}%
\subfloat[]{%
  \includegraphics[width=0.24\columnwidth]{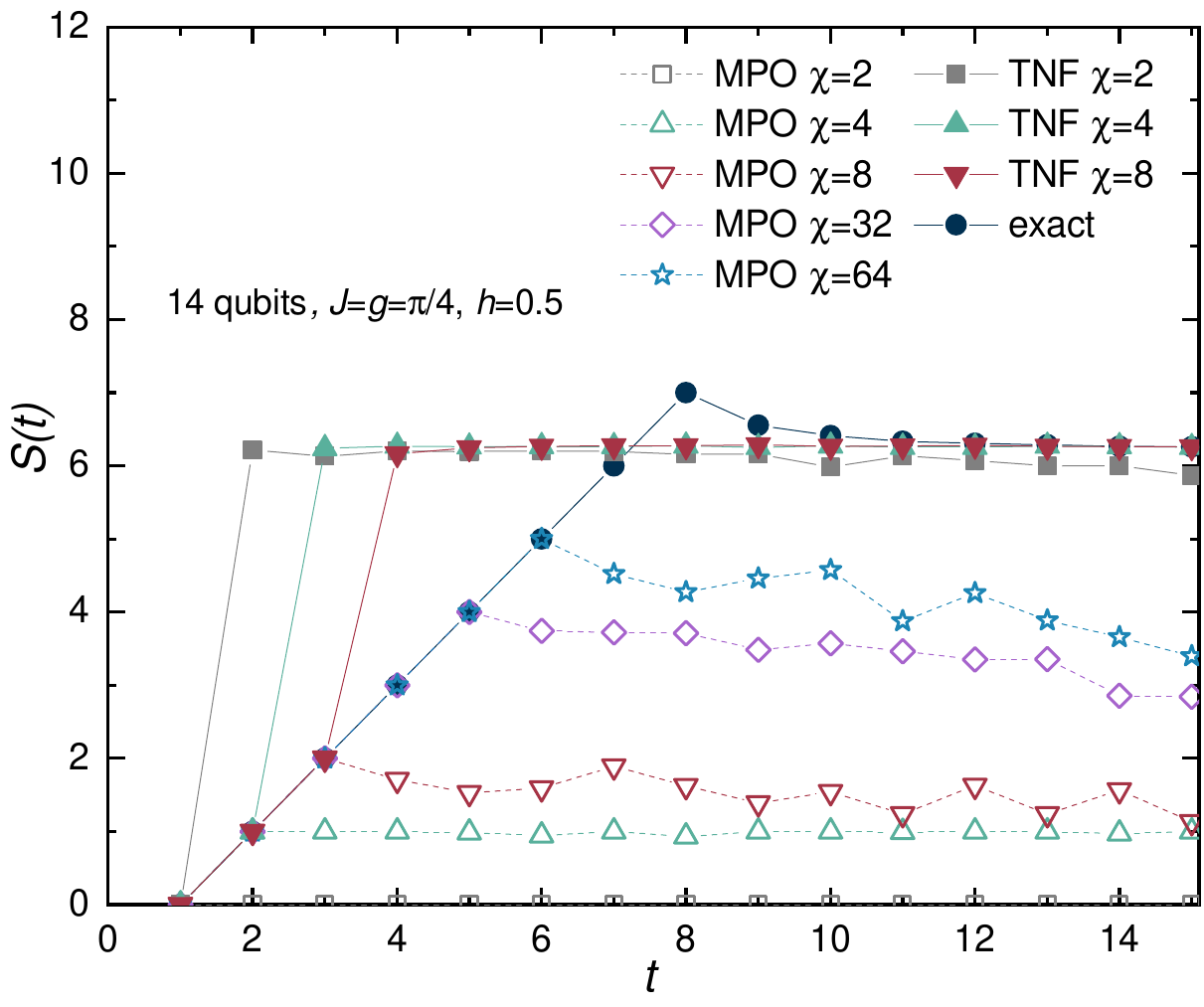}
}\hspace*{\fill}   

\caption{\label{fig:inverse} Half-chain entanglement entropy dynamics for $L = 10$ (a)-(d) and $L = 14$ (e)-(h) kicked Ising chain, obtained from different approaches. Inverse-time contraction is used for the TN function. For (a),(c),(e) and (g), conventional MPS-MPO contraction along the time direction is used (labeled as MPS in legends). For (b),(d),(f) and (h),  MPO-MPO contraction along the inverse time direction is used (labeled as MPO in legends).} 
\end{figure}

In Fig.~\ref{fig:inverse}, we show the numerical results for the entanglement entropy dynamics of the 1D kicked Ising model for two different parameter sets for $L=10$ and $L=14$ spin chains, where the tensor network functions are contracted along the inverse-time direction and the conventional MPS results are also shown for comparison. In addition, we show results from an inverse-time MPO-MPO contraction (i.e Heisenberg-style  evolution) for comparison. In the latter case, the isometries are not amplitude dependent, as the amplitude is evaluated only after the entire tensor network is contracted (in the inverse time direction) into an MPS.

One can see that  with the same $\chi$, the tensor network function contracted in the inverse direction captures more entanglement than the conventional MPS, as well as the Heisenberg-style MPS from MPO-MPO contraction. However, compared to the transverse contraction, inverse-time contraction has a lower accuracy at the early steps of the Floquet dynamics, showing a convergence of the entanglement entropy ahead of the true saturation time for the maximally chaotic kicked Ising dynamics.

\section{S-3. Additional discussion of the tensor network function representation of neural network computational graphs}

\subsection{A. Classical binary arithmetic circuits}

In our first construction of the tensor network function representation of a feed forward neural network, we use classical binary arithmetic using logic and copy gates. Here we describe in more detail this binary arithmetic and the building blocks (in terms of the tensors) of polynomial functions. {(We can  use polynomial functions to efficiently represent activation functions in a neural network under mild conditions).} {Note that we have chosen the construction below for simplicity and this is not the only binary logic construction of arithmetic: for more examples, we refer to~\cite{kumar2010digital,lancaster2001excel,mano2017digital,1672371}.}

\begin{figure}[htb]
    \centering

\subfloat[\label{fig:binary_add1}]{%
  \includegraphics[width=0.13\columnwidth]{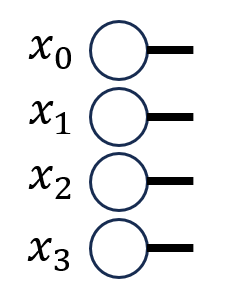}%
}\hspace*{\fill}%
\subfloat[ \label{fig:binary_add2}]{%
  \includegraphics[width=0.25\columnwidth]{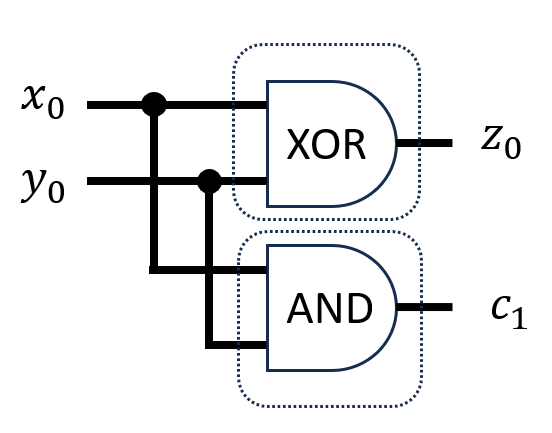}%
}\hspace*{\fill}%
\subfloat[ \label{fig:binary_add3}]{%
  \includegraphics[width=0.35\columnwidth]{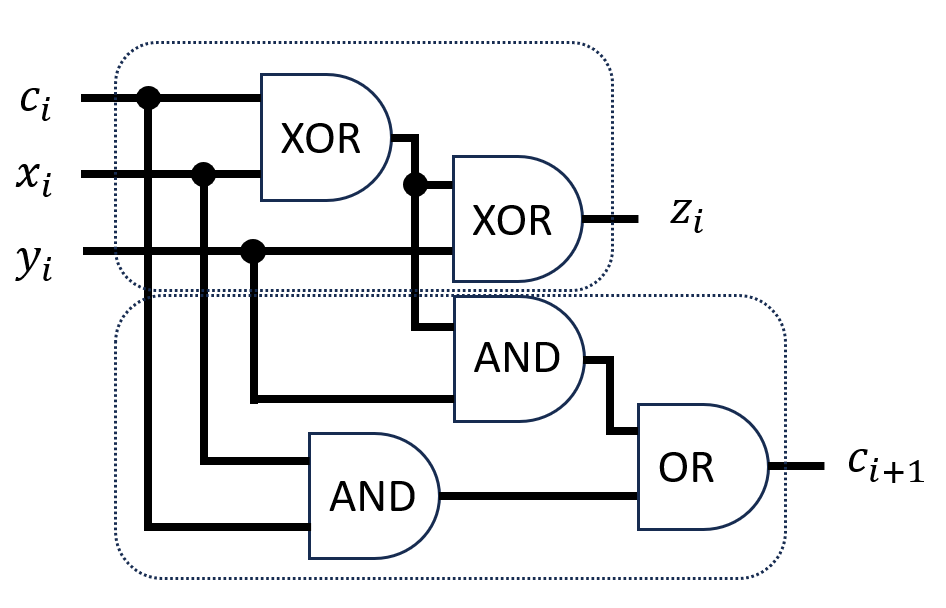}%
}\hspace*{\fill}%
\subfloat[ \label{fig:binary_add4}]{%
  \includegraphics[width=0.25\columnwidth]{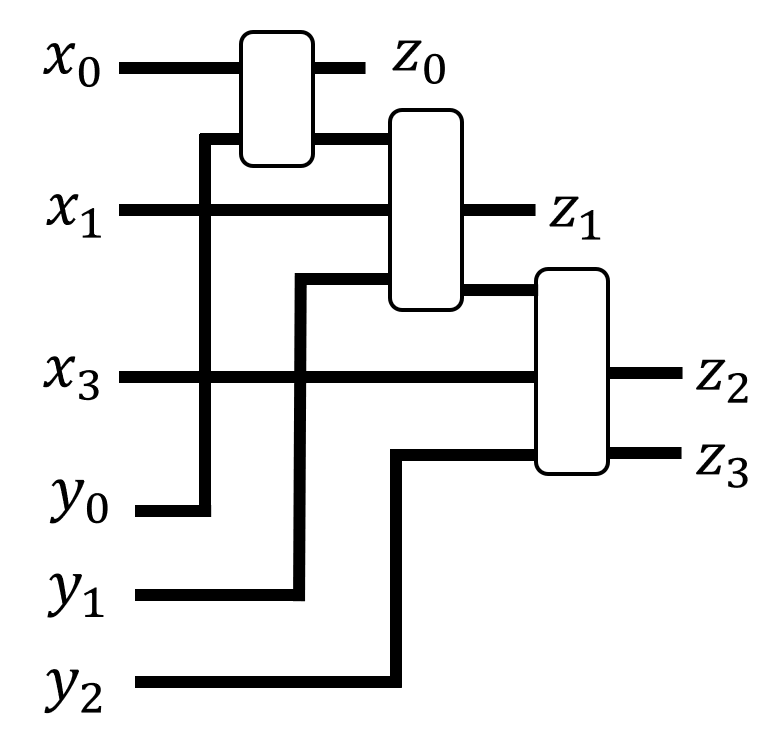}%
}\hspace*{\fill}%

    \caption{(a) Representing a bit string as a product of length-2 vectors. (b) Diagram for adding 2 bits. (c) Diagram for adding 3 bits. (d) Diagram for $x+y=z$ where $x$ and $y$ each has 3 digits.}
    \label{fig:binary_add}
\end{figure}

The $n$-bit binary representation of an integer 
\begin{align}
    x=\sum_{i=0}^nx_i2^i,\quad x_i\in\{0,1\}
\end{align}
is a product of length-2 vectors as shown in Fig.~\ref{fig:binary_add1}, where each vector 
\begin{align}\label{eqn:bit}
    (x_i)_j=\delta_{x_i,j};
\end{align}
floating point numbers can be represented similarly. Consider addition
\begin{align}
    z=x+y=\sum_{i=0}^n(x_i+y_i)2^i.
\end{align}
The tensor network function to  compute the 0-th digit 
\begin{align}
    z_0=(x_0+y_0)\text{ mod }2
\end{align}
is shown in the upper dotted box of Fig.~\ref{fig:binary_add2}, where the XOR tensor of shape (2,2,2) has zero entries except at
\begin{align}
    (\text{XOR})_{0,0,0}=(\text{XOR})_{1,0,1}=(\text{XOR})_{0,1,1}=(\text{XOR})_{1,1,0}=1.
\end{align}
The contribution carried to the 1st digit
\begin{align}
    c_1=\text{floor}((x_0+y_0)/2)
\end{align}
is shown in the lower dotted box of Fig.~\ref{fig:binary_add2} where the AND tensor of shape (2,2,2) has zero entries except at
\begin{align}\label{eq:and}
    (\text{AND})_{0,0,0}=(\text{AND})_{1,0,0}=(\text{AND})_{0,1,0}=(\text{AND})_{1,1,1}=1.
\end{align}
Fig.~\ref{fig:binary_add2} is also known as a "half adder" in an electric circuit \cite{kumar2010digital,lancaster2001excel,mano2017digital}. For any subsequent digit $i$, the diagram to compute
\begin{align}
    z_i=(c_i+x_i+y_i)\text{ mod }2
\end{align}
is shown in the upper dotted box of Fig.~\ref{fig:binary_add3}. The contribution carried over to the next digit 
\begin{align}
    c_{i+1}=\text{floor}((c_i+x_i+y_i)/2)
\end{align}
is shown in the lower dotted box of Fig.~\ref{fig:binary_add3}, where the OR tensor of shape (2,2,2) has zero entries except at
\begin{align}
    (\text{OR})_{0,0,0}=(\text{OR})_{1,0,1}=(\text{OR})_{0,1,1}=(\text{OR})_{1,1,1}=1. 
\end{align}
Fig.~\ref{fig:binary_add3} is also known as a "full adder" in an electric circuit\cite{kumar2010digital,lancaster2001excel,mano2017digital,1672371}. In Fig.~\ref{fig:binary_add4} we show the full circuit representation for $x+y=z$ with each of $x$ and $y$ containing 3 digits, and where the boxes with 2 and 3 input legs represent the tensor network subgraph in Fig.~\ref{fig:binary_add2} and Fig.~\ref{fig:binary_add3} respectively. 

\begin{figure}[htb]
    \centering
    \includegraphics[width=0.7\linewidth]{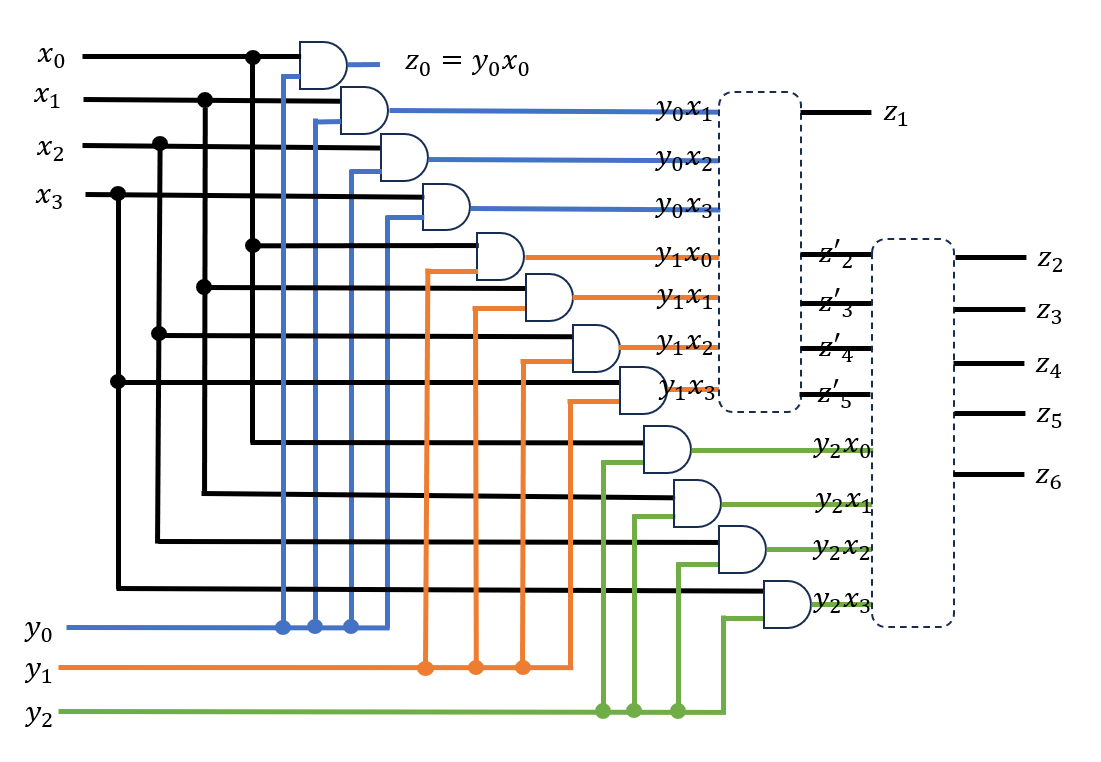}
\caption{Tensor network for computing $z=xy$ with $x=2^3x_3+2^2x_2+2x_1+x_0$ and $y=2^2y_2+2y_1+y_0$. Partial products $y_0x$, $y_1x$ and $y_2x$ are represented in blue, orange and green. The dashed boxes represent the subgraph for adding 2 bit strings as shown in Fig.~\ref{fig:binary_add4}. }
    \label{fig:binary_multiply}
\end{figure}

Multiplication 
\begin{align}
    z=x\times y=\left(\sum_{i=0}^mx_i2^i\right)\left(\sum_{j=0}^ny_i2^j\right)%=\sum_{i=0}^m\sum_{j=0}^nx_iy_j2^{i+j}
\end{align}
in the binary arithmetic circuit representation can be built from binary long multiplication. Here we illustrate this with an example for $m=3$, $n=2$, where long multiplication 
\begin{align}
\begin{array}{ccccccc}
& &  & x_3 & x_2 & x_1 & x_0 \\
& & \times &  & y_2 & y_1 & y_0 \\
\hline 
& &  & y_0x_3 & y_0x_2 & y_0x_1 & y_0x_0 \\
& + & y_1x_3 & y_1x_2 & y_1x_1 & y_1x_0 & \\
+ & y_2x_3 & y_2x_2 & y_2x_1 & y_2x_0 & & 
    \end{array}
\end{align}
first forms partial products shown in each row under the line and then sums the shifted partial products. The corresponding tensor network is shown in Fig.~\ref{fig:binary_multiply}, where all 3-legged tensors are the AND tensors given in Eq.~\ref{eq:and}, and partial products of $y_0x$, $y_1x$ and $y_2x$ are represented by blue, orange and green respectively. The summation of partial products is done sequentially using the tensor network subgraph shown in Fig.~\ref{fig:binary_add4} represented as the dashed boxes in Fig.~\ref{fig:binary_multiply}. In digital circuits, various more efficient alternatives exist~\cite{1672241,1052564,8228275}.  

\begin{figure}[htb]
    \centering
    \includegraphics[width=0.5\linewidth]{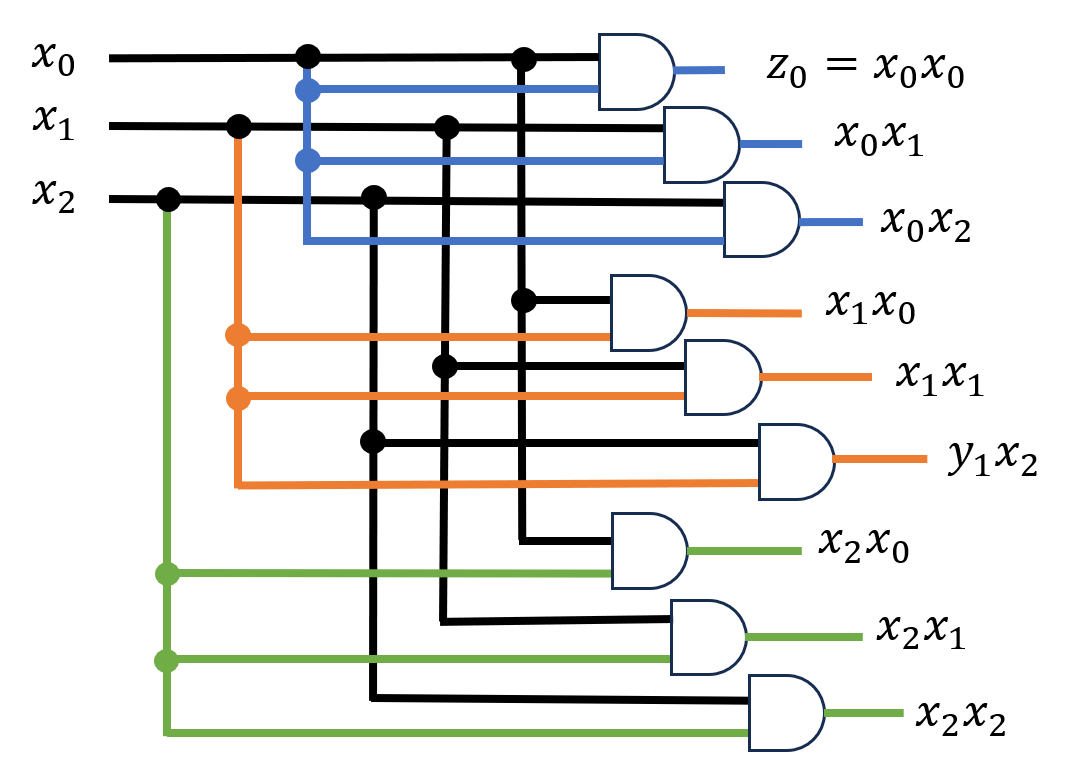}
  \caption{Tensor network for computing $z=x^2$ with $x=2^2x_2+2x_1+x_0$. Partial products $x_0x$, $x_1x$ and $x_2x$ are represented in blue, orange and green.}
    \label{fig:binary_square}
\end{figure}

Finally, we comment on the copying of data. In the binary arithmetic circuit, the information contained in a bit can be fully copied by the $\delta$ tensor which has zero entries except at 
\begin{align}\label{eqn:delta_small}
    \delta_{0,0,0}=\delta_{1,1,1}=1
\end{align}
and is represented as the black dot in Fig.~\ref{fig:binary_add2}, Fig.~\ref{fig:binary_add3} and Fig.~\ref{fig:binary_multiply}. This is because a bit represents its value by the index associated with the nonzero amplitude, which is 1, but the nonzero amplitude itself contains no additional information. Therefore, let $x$ represent a bit, then 
\begin{align}\label{eqn:bit_delta}
\sum_ix_i\delta_{ijk}=\sum_i\delta_{x,i}\delta_{ijk}=\delta_{xjk}=\delta_{x,j}\delta_{x,k}=x_jx_k
\end{align}
where we use the representation given in Eq.~\ref{eqn:bit}. To illustrate this point, we show the tensor network subgraph for computing $z=x^2$ in Fig.~\ref{fig:binary_square} upto the formation of all partial products, where only one set of $x$-bits is needed as input. 

Furthermore, we note that contracting a bit with the $\delta$ tensor results in a product of 2 vectors, as in Eq.~\ref{eqn:bit_delta}. Therefore, for a given pair of bit inputs, the output of Fig.~\ref{fig:binary_add2} is a product; and similarly for Fig.~\ref{fig:binary_add3}. Hence for a given bit string input, Fig.~\ref{fig:binary_add4} and thus Fig.~\ref{fig:binary_multiply} can be easily contracted.

\subsection{B. Amplitude arithmetic circuit construction of tensor network functions}
\begin{figure}[htb]
    \centering
        
\subfloat[\label{fig:amplitude1}]{%
  \includegraphics[width=0.1\columnwidth]{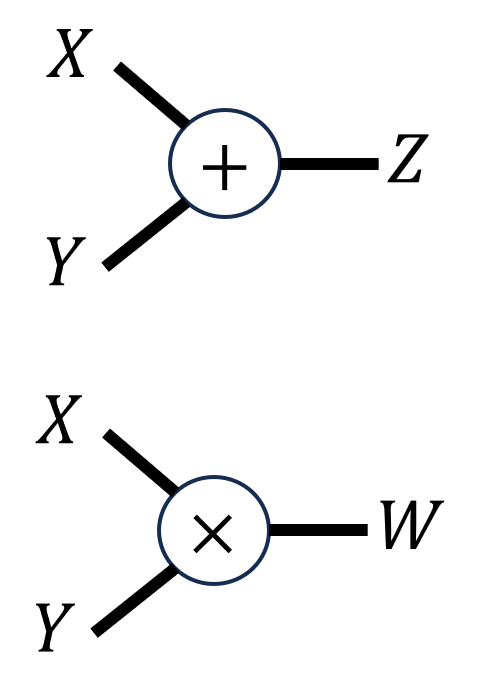}%
}\hspace*{\fill}%
\subfloat[\label{fig:amplitude2}]{%
  \includegraphics[width=0.2\columnwidth]{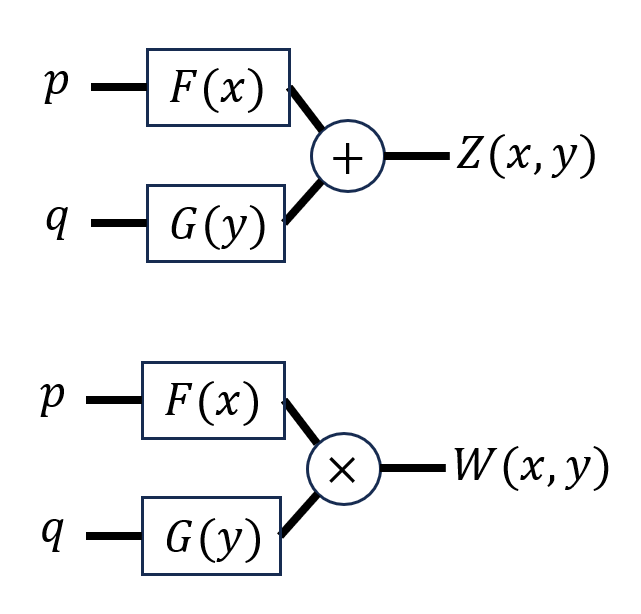}%
}\hspace*{\fill}%
\subfloat[\label{fig:amplitude3}]{%
  \includegraphics[width=0.2\columnwidth]{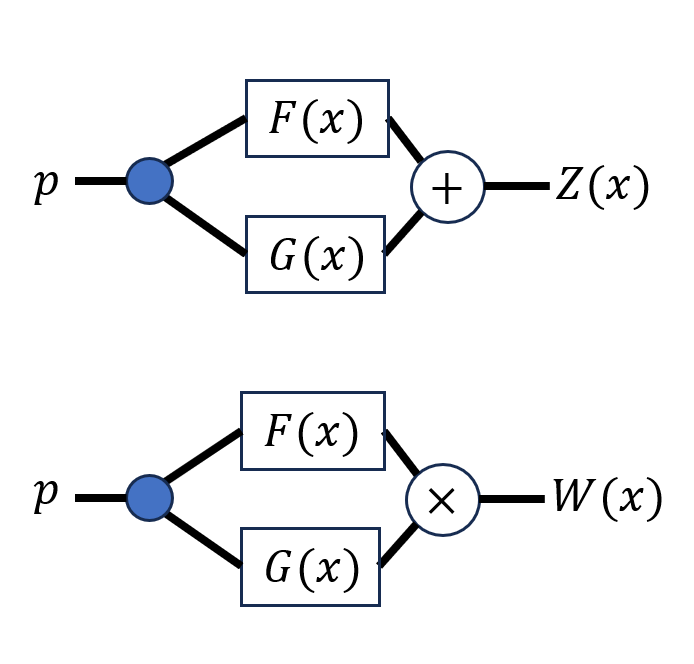}%
}\hspace*{\fill}%

    \caption{ (a) Tensor diagrams for addition $z=x+y$ and for multiplication $w=xy$. (b) Tensor network diagrams for addition and multiplication of functions $z(x[p],y[q])=f(x[p])+g(y[q])$ and $z(x[p],y[q])=f(x[p])g(y[q])$ where $p,q$ index the discretized variables $x,y$ respectively. (c) Tensor network diagrams for addition and multiplication of functions $z(x[p])=f(x[p])+g(x[p])$ and $z(x[p])=f(x[p])g(x[p])$. }
    \label{fig:amplitude}

\end{figure}

The arithmetic circuit construction in Ref.~\cite{peng2023} represents a floating point number $x$ as a length-2 vector
\begin{align}
    X_i=x^i,\quad i=0,1
\end{align}
where superscript $i$ denotes raising to $i$ the power. The diagrams for addition $z=x+y$ and for multiplication $w=xy$ are shown in Fig.~\ref{fig:amplitude1}, where the tensors $(+)$ and $(\times)$ have shape $(2,2,2)$ and entries 
\begin{align}    (+)_{ijk}=\delta_{i+j,k},\quad(\times)_{ijk}=\delta_{ijk}
\end{align}
for $i,j,k=0,1$ such that 
\begin{align}
&Z_k=\sum_{ij}X_iY_j(+)_{ijk}=\sum_{ij}x^iy^j\delta_{i+j,k}=(x+y)^k\\
&W_k=\sum_{ij}X_iY_j(\times)_{ijk}=\sum_{ij}x^iy^j\delta_{ijk}=(xy)^k.
\end{align}

{The above representation allows us to efficiently compose more complicated functions using the addition and multiplication tensors and floating point numbers, but special consideration must given to copying data (i.e. to construct a non-linear dependence).} 
To illustrate the difference in copying data from the case of classical binary circuits, we consider the addition of single variable functions $z(x,y)=f(x)+g(y)$ shown in the upper diagram in Fig.~\ref{fig:amplitude2}; 
multiplication is similar by replacing the $(+)$ tensor with $(\times)$. In the following we will use $p,q,r,...$ to denote tensor legs associated with discretized variables, and $i,j,k,...$ to denote the size-2 leg contracted with the $(+)$ and $(\times)$ tensors for performing arithmetic. The addition (multiplication) of single variable functions of the same variable $z(x)=f(x)+g(x)$($w(x)=f(x)g(x)$) is shown in Fig.~\ref{fig:amplitude3}, where the blue dot represents the copy tensor on the variable leg $\delta_{pqr}$. In particular, given tensor $F_{p,i}$ representing the discretized function $f(x)$, the tensor network representation of $w(x)=f(x)^2$ requires 2 copies of $F_{p,i}$ with the variable legs constrained to be the same by the $\delta_{pqr}$ tensor. This is to be contrasted with the binary representation, where a single copy of a bit string $x$ is needed to construct the tensor network representation of $x^2$. In general, as demonstrated in the previous section, copying of data in the binary representation is done using the shape $(2,2,2)$ $\delta_{ijk}$ tensor in Eq.~\ref{eqn:delta_small} from a \emph{single} copy of the bit string. In contrast, copying of data in the amplitude representation requires additional actual copies of the same tensor. 

In principle, the requirement of actual copies of tensors in the amplitude representation can lead to exponential growth in the number of tensors in the full tensor network function representation of a computational graph, for instance, for computing an amplitude for a given input in a deep neural network. However, this formal exponential full tensor network representation is never necessary in actual computation. As in the case of the binary circuit representation, contraction of the amplitude circuit representation can proceed by trivially following the computational graph and storing intermediates as needed. So long as we augment the tensor network computational model with the ability to store intermediates in a ``memory'', we can replace copied subgraphs of the tensor network function by their evaluated value when it is available. Therefore, contraction of the full circuit representation can be done without forming the full tensor network, and the contraction has the same cost as the computational graph that it trivially follows. 

\end{document}